\documentclass[review]{elsarticle}

\usepackage{lineno,hyperref}
\modulolinenumbers[5]

\usepackage[misc]{ifsym}
\usepackage{amssymb}
\usepackage{amsmath}

\usepackage{todonotes}

\usepackage{soul}

\usepackage{graphicx}

\usepackage[caption=false]{subfig}
\usepackage{float}

\usepackage{manyfoot}

\DeclareNewFootnote{A}
\DeclareNewFootnote{B}

\let\footnoteSym\footnoteA

\graphicspath{ {./Figures/} }
\journal{Computer Methods and Programs in Biomedicine}

\biboptions{sort&compress}

\begin{document}

\begin{titlepage}
	\clearpage\thispagestyle{empty}
		
\begin{center}
\begin{large}\textbf{3D deformable registration of longitudinal abdominopelvic \\CT images using unsupervised deep learning}\end{large}
\end{center}

\textbf{\\Short abstract:}
This study investigates the use of the unsupervised deep learning framework VoxelMorph for deformable registration of longitudinal abdominopelvic CT images acquired in patients with bone metastases from breast cancer.
The CT images were refined prior to registration by automatically removing the CT table and all other extra-corporeal components.
To improve the learning capabilities of VoxelMorph when only a limited amount of training data is available, a novel incremental training strategy is proposed based on simulated deformations of consecutive CT images.
In a 4-fold cross-validation scheme, the incremental training strategy achieved significantly better registration performance compared to training on a single volume.
Although our deformable image registration method did not outperform iterative registration using NiftyReg (considered as a benchmark) in terms of registration quality, the registrations were approximately $300$ times faster.
This study showed the feasibility of deep learning based deformable registration of longitudinal abdominopelvic CT images \textit{via} a novel incremental training strategy based on simulated deformations.


\newpage

\noindent Maureen van Eijnatten\footnoteSym[2]{These authors contributed equally}\textsuperscript{,}\footnoteSym[1]{Corresponding author.\\  \textit{Address}: Medical Image Analysis Group, Department of Biomedical Engineering, Eindhoven
University of Technology, P.O. Box 513, 5600 MB Eindhoven, The Netherlands.}, \textit{E-mail}: \texttt{mve@cwi.nl; m.a.j.m.v.eijnatten@tue.nl}\\
\begin{scriptsize}
Centrum Wiskunde \& Informatica, 1098 XG Amsterdam, The Netherlands\\
Medical Image Analysis Group, Department of Biomedical Engineering, Eindhoven
University of Technology, 5600 MB Eindhoven, The Netherlands
\end{scriptsize}

\noindent Leonardo Rundo$^{\dagger}$,
\textit{E-mail}: \texttt{lr495@cam.ac.uk}\\
\begin{scriptsize}
Department of Radiology, University of Cambridge, CB2 0QQ Cambridge, UK\\
Cancer Research UK Cambridge Centre, University of Cambridge, CB2 0RE Cambridge, UK
\end{scriptsize}

\noindent K. Joost Batenburg, \textit{E-mail}: \texttt{k.j.batenburg@cwi.nl}\\
\begin{scriptsize}
Centrum Wiskunde \& Informatica, 1098 XG Amsterdam, The Netherlands\\
Leiden Institute of Advanced Computer Science, Leiden University, 2300 RA Leiden, The Netherlands
\end{scriptsize}

\noindent Felix Lucka, \textit{E-mail}: \texttt{felix.lucka@cwi.nl}\\
\begin{scriptsize}
Centrum Wiskunde \& Informatica, 1098 XG Amsterdam, The Netherlands\\
Centre for Medical Image Computing, University College London, WC1E 6BT London, United Kingdom
\end{scriptsize}

\noindent Emma Beddowes, \textit{E-mail}: \texttt{emma.beddowes@cruk.cam.ac.uk}\\
\begin{scriptsize}
Cancer Research UK Cambridge Centre, University of Cambridge, CB2 0RE Cambridge, UK\\
Cancer Research UK Cambridge Institute, CB2 0RE University of Cambridge, Cambridge, United Kingdom\\
Department of Oncology, Addenbrooke's Hospital, Cambridge University Hospitals National Health Service (NHS) Foundation Trust, CB2 0QQ Cambridge, United Kingdom
\end{scriptsize}

\noindent Carlos Caldas, \textit{E-mail}: \texttt{carlos.caldas@cruk.cam.ac.uk}\\
\begin{scriptsize}
Cancer Research UK Cambridge Centre, University of Cambridge, CB2 0RE Cambridge, UK\\
Cancer Research UK Cambridge Institute, CB2 0RE University of Cambridge, Cambridge, United Kingdom\\
Department of Oncology, Addenbrooke's Hospital, Cambridge University Hospitals National Health Service (NHS) Foundation Trust, CB2 0QQ Cambridge, United Kingdom
\end{scriptsize}

\noindent Ferdia A. Gallagher, \textit{E-mail}: \texttt{fag1000@cam.ac.uk}\\
\begin{scriptsize}
Department of Radiology, University of Cambridge, CB2 0QQ Cambridge, UK\\
Cancer Research UK Cambridge Centre, University of Cambridge, CB2 0RE Cambridge, UK
\end{scriptsize}

\noindent Evis Sala, \textit{E-mail}: \texttt{es220@cam.ac.uk}\\
\begin{scriptsize}
Department of Radiology, University of Cambridge, CB2 0QQ Cambridge, UK\\
Cancer Research UK Cambridge Centre, University of Cambridge, CB2 0RE Cambridge, UK
\end{scriptsize}

\noindent Carola-Bibiane Sch\"{o}nlieb\footnoteSym[3]{These authors equally co-supervised the work.}, \textit{E-mail}: \texttt{cbs31@cam.ac.uk}\\
\begin{scriptsize}
Department of Applied Mathematics and Theoretical Physics, University of Cambridge, CB3 0WA Cambridge, United Kingdom
\end{scriptsize}

\noindent Ramona Woitek$^{\ddagger}$, \textit{E-mail}: \texttt{rw585@cam.ac.uk}\\
\begin{scriptsize}
Department of Radiology, University of Cambridge, CB2 0QQ Cambridge, UK\\
Cancer Research UK Cambridge Centre, University of Cambridge, CB2 0RE Cambridge, UK\\
Department of Biomedical Imaging and Image-guided Therapy, Medical University Vienna, 1090 Vienna, Austria
\end{scriptsize}

\newpage
\end{titlepage}

\begin{frontmatter}

\title{3D deformable registration of longitudinal \\abdominopelvic CT images using \\unsupervised deep learning}

\author[CWI,TUe]{Maureen van Eijnatten\corref{correspondingauthor}\footnoteSym[2]{These authors contributed equally.}\textsuperscript{,}}
\cortext[correspondingauthor]{Corresponding author.\\ \textit{E-mail address:} \texttt{mve@cwi.nl; m.a.j.m.v.eijnatten@tue.nl}}

\author[Radiol,CRUKCC]{Leonardo Rundo$^{\dagger,}$}

\author[CWI,LU]{K. Joost Batenburg}

\author[CWI,UCL]{Felix Lucka}

\author[CRUKCC,CRUKCI,Oncol]{Emma Beddowes}

\author[CRUKCC,CRUKCI,Oncol]{Carlos Caldas}

\author[Radiol,CRUKCC]{Ferdia A. Gallagher}

\author[Radiol,CRUKCC]{Evis Sala}

\author[DAMTP]{Carola-Bibiane Sch\"{o}nlieb\footnoteSym[3]{These authors equally co-supervised the work.}\textsuperscript{,}}

\author[Radiol,CRUKCC,Vienna]{Ramona Woitek$^{\ddag,}$}

\address[CWI]{Centrum Wiskunde \& Informatica, 1098 XG Amsterdam, The Netherlands}
\address[TUe]{Medical Image Analysis Group, Department of Biomedical Engineering,\\Eindhoven
University of Technology, 5600 MB Eindhoven, The Netherlands}
\address[Radiol]{Department of Radiology, University of Cambridge, CB2 0QQ Cambridge, United Kingdom}
\address[CRUKCC]{Cancer Research UK Cambridge Centre, University of Cambridge,\\CB2 0RE Cambridge, United Kingdom}
\address[LU]{Mathematical Institute, Leiden University, 2300 RA Leiden, The Netherlands}
\address[UCL]{Centre for Medical Image Computing, University College London, WC1E 6BT London, United Kingdom}
\address[CRUKCI]{Cancer Research UK Cambridge Institute, University of Cambridge,\\ CB2 0RE Cambridge, United Kingdom}
\address[Oncol]{Department of Oncology, Addenbrooke's Hospital, Cambridge University Hospitals National Health Service (NHS) Foundation Trust, CB2 0QQ Cambridge, United Kingdom}
\address[DAMTP]{Department of Applied Mathematics and Theoretical Physics, University of Cambridge, CB3 0WA Cambridge, United Kingdom}
\address[Vienna]{Department of Biomedical Imaging and Image-guided Therapy,\\Medical University Vienna, 1090 Vienna, Austria}

\begin{abstract}

\noindent\textit{Background and Objectives:}
Deep learning is being increasingly used for deformable image registration and unsupervised approaches, in particular, have shown great potential.
However, the registration of abdominopelvic Computed Tomography (CT) images remains challenging due to the larger displacements compared to those in brain or prostate Magnetic Resonance Imaging datasets that are typically considered as benchmarks.
In this study, we investigate the use of the commonly used unsupervised deep learning framework VoxelMorph for the registration of a longitudinal abdominopelvic CT dataset acquired in patients with bone metastases from breast cancer.

\noindent\textit{Methods:}
As a pre-processing step, the abdominopelvic CT images were refined by automatically removing the CT table and all other extra-corporeal components.
To improve the learning capabilities of the VoxelMorph framework when only a limited amount of training data is available, a novel incremental training strategy is proposed based on simulated deformations of consecutive CT images in the longitudinal dataset.
This devised training strategy was compared against training on simulated deformations of a single CT volume.
A widely used software toolbox for deformable image registration called NiftyReg was used as a benchmark.
The evaluations were performed by calculating the Dice Similarity Coefficient (DSC) between manual vertebrae segmentations and the Structural Similarity Index (SSIM).

\noindent\textit{Results:}
The CT table removal procedure allowed both VoxelMorph and NiftyReg to achieve significantly better registration performance.
In a 4-fold cross-validation scheme, the incremental training strategy resulted in better registration performance compared to training on a single volume, with a mean DSC of $0.929 \pm 0.037$ and $0.883 \pm 0.033$, and a mean SSIM of $0.984 \pm 0.009$ and $0.969 \pm 0.007$, respectively.
Although our deformable image registration method did not outperform NiftyReg in terms of DSC ($0.988 \pm 0.003$) or SSIM ($0.995 \pm 0.002$), the registrations were approximately $300$ times faster.

\noindent\textit{Conclusions:}
This study showed the feasibility of deep learning based deformable registration of longitudinal abdominopelvic CT images \textit{via} a novel incremental training strategy based on simulated deformations.
\end{abstract}


\end{frontmatter}


\section{Introduction}
\label{sec:Intro}

Deformable medical image registration problems can be solved by optimizing an objective function defined on the space of transformation parameters \cite{pluim2000}.
Traditional optimization-based methods typically achieve accurate registration results but suffer from being computationally expensive, especially in the case of deformable transformations of high-resolution, three-dimensional (3D) images.
Deep learning based registration methods, however, can perform registration in a single-shot, which is considerably faster than using iterative methods \cite{sokooti2017}.
Due to the recent successes of deep learning for a wide variety of medical image analysis tasks \cite{litjens2017survey}, and the advances in Graphics Processing Unit (GPU) computing that have enabled the training of increasingly large three-dimensional (3D) networks \cite{shen2017}, the number of studies using deep learning for medical image registration has increased considerably since 2016 \cite{haskins2020}.

Although deep learning could have a major impact on the field of medical image registration, there is still a gap between proof-of-concept technical feasibility studies and the application of these methods to ``real-world'' medical imaging scenarios.
It remains unclear to which extent deep learning is suited for challenging co-registration tasks with large inter- and intra-patient variations and potential outliers or foreign objects in the Volume of Interest (VOI).
Moreover, deep learning based methods typically require large amounts---i.e., thousands---of well prepared, annotated 3D training images that are rarely available in clinical settings \cite{han2019CIKM}. 

The present study focuses on the registration of abdominopelvic CT images since these are widely acknowledged to be difficult to register \cite{xu2016}.
In abdominopelvic imaging, the conservation-of-mass assumption is typically not valid and, although local-affine diffeomorphic demons have been used in abdominal CT images \cite{freiman2011}, the transformation is typically not a diffeomorphism.
For instance, bladder-filling or bowel peristalsis in the abdomen may vary between images. 
More specifically, we consider a \textit{longitudinal} abdominopelvic CT dataset that comprises several images of each patient acquired at distinct time points.
From a clinical perspective, deformable image registration of longitudinal datasets is a necessary step toward automated, quantitative, and objective treatment-response assessment~\cite{yankeelov2016,blackledge2014,reischauer2018}.

The aim of this study is to explore and quantify the applicability of one of the most used unsupervised single-shot deep learning frameworks (VoxelMorph~\cite{balakrishnan2019}) for deformable registration of longitudinal abdominopelvic CT images.
We assessed the maximum displacements that can be learned by the VoxelMorph framework and the impact of extra-corporeal structures, such as the CT table, clothing and prostheses on the registration performance.
In addition, the VoxelMorph framework was compared against iterative registration using the NiftyReg~\cite{modat2010} toolbox that was selected because of its excellent performance on abdominal CT images in a comparative study \cite{lee2015evaluation}.
 
The novelties of this work are:
\begin{itemize}
    \item demonstrating the impact of removing extracorporeal structures before deformable image registration;
    \item using simulated deformations to characterize the limitations of the VoxelMorph framework for the deformable registration of abdominopelvic CT images;
    \item introducing a novel incremental training strategy tailored to longitudinal datasets that enables deep learning based image registration when limited amounts of training data are available.
\end{itemize}

This paper is structured as follows.
Section~\ref{sec:relatedWork} outlines the background of medical image registration, with a particular focus on deep learning based methods.
Section~\ref{sec:MatMet} presents the characteristics of our longitudinal abdominopelvic CT dataset, as well as the deformable registration framework, the proposed incremental training strategy, and the evaluation metrics used in this study. 
Section~\ref{sec:results} describes the experimental results.
Finally, Section~\ref{sec:discussion} provides a discussion and concluding remarks.

\section{Related work}
\label{sec:relatedWork}

This section introduces the basic concepts of medical image registration and provides a comprehensive overview about the state-of-the-art of deformable registration using deep learning.

\subsection{Medical image registration}
\label{sec:medimreg}

Medical image registration methods aim to estimate the best solution in the parameter space $\Omega \subset \mathbb{R}^N$ which corresponds to the set of potential transformations used to align the images, where $N$ is the number of dimensions.
Typically, $N \in \{ 2,3 \}$ in biomedical imaging.
Each point in $\Omega$ corresponds to a different estimate of the transformation that maps a moving image to a fixed image (target).
This transformation can be either parametric, i.e., can be parameterized by a small number of variables (e.g., six in case of a 3D rigid-body transformation or twelve for an 3D affine transformation), or non-parametric, i.e., in the case that we seek the displacement of every image element.
For most organs in the human body, particularly in the abdomen, many degrees of freedom are necessary to deal with non-linear or local soft-tissue deformations.
In global deformable transformation, the number of parameters encoded in a Displacement Vector Field (DVF) $\phi$ is typically large, e.g., several thousands.
Therefore, two-step intensity-based registration approaches are commonly employed in which the first step is a global affine registration and the second step is a local deformable registration using for example B-splines~\cite{klein2007evaluation}.

Traditional medical image registration methods often use iterative optimization techniques based on gradient descent to find the optimal transformation \cite{pluim2000,klein2007evaluation,bernon2001}.
Deformable registration can be performed using demons \cite{thirion1998}, typically based on diffeomorphic transformations parameterized by stationary velocity fields \cite{vercauteren2009}.
In addition, global optimization techniques that leverage evolutionary algorithms \cite{klein2007evaluation} and swarm intelligence meta-heuristics  can be useful to avoid local minima \cite{rundo2016SSCI}.
Several off-the-shelf, open-source toolboxes are available for both parametric and non-parametric image registration in biomedical research, such as: \texttt{elastix}~\cite{klein2009elastix}, NiftyReg~\cite{modat2010}, Advanced Normalization Tools (ANTs)~\cite{tustison2014}, and Flexible Algorithms for Image Registration (FAIR) \cite{modersitzki2009FAIR}.

\subsection{Deep learning based registration}
\label{sec:deepLearning}

Since 2013, the scientific community has shown an increasing interest in medical image registration based on deep learning \cite{haskins2020}.
Early unsupervised deep learning based registration approaches leveraged stacked convolutional neural networks (CNNs) or autoencoders to learn the hierarchical representations for patches~\cite{wu2013,wu2016}.

Fully-supervised methods, such as in~\cite{Cheng2018}, have focused on learning a similarity metric for multi-modal CT-MRI brain registration according to the patch-based correspondence.
Another supervised method based on the Large Deformation Diffeomorphic Metric Mapping (LDDMM) model called Quicksilver was proposed in \cite{yang2017} and tested on brain MRI scans.
In this context, Eppenhof \textit{et al.}~\cite{eppenhof2019TMI} introduced the simulation of ground truth deformable transformations to be employed during training to overcome the need for manual annotations in the case of a pulmonary CT dataset.
Very recently, in~\cite{Ha2019MIDL}, a graph CNN was used to estimate global key-point locations and regress the relative displacement vectors for sparse correspondences.

Alternatively, several studies have focused on weakly-supervised learning.
For example, Hu \textit{et al.} \cite{hu2018} proposed a weakly-supervised framework for 3D multimodal registration.
This end-to-end CNN approach aimed to predict displacement fields to align multiple labeled corresponding structures for individual image pairs during the training, while only unlabeled image pairs were used as network input for inference.
Recently, generative deep models have also been applied to unsupervised deformable registration.
Generative Adversarial Networks (GANs) can be exploited as an adversarial learning approach to constrain CNN training for deformable image registration, such as in \cite{hu2018adversarial} and \cite{yan2018adversarial}.
In \cite{tanner2018}, spatial correspondence problems due to the different acquisition conditions (e.g., inhale-exhale states) in MRI-CT deformable registration, led to changes synthesized by the adversarial learning, which were addressed by reducing the size of the discriminator's receptive fields.
In addition, Krebs \textit{et al.} ~\cite{krebs2019} proposed a probabilistic model for diffeomorphic registration that leverages Conditional Variational Autoencoders.

The current trend in deep learning based medical image registration is moving towards unsupervised learning \cite{haskins2020}.
The CNN architecture proposed in~\cite{sokooti2017}, called RegNet---different from existing work---directly estimates the displacement vector field from a pair of input images; it integrates image content at multiple scales by means of a dual path, allowing for contextual information.
Traditional registration methods optimize an objective function independently for each pair of images, which is time-consuming for large-scale datasets.
To this end, the differentiable Spatial Transformer Layer (STL) has been introduced that enables CNNs to perform global parametric image alignment without requiring supervised labels \cite{jaderberg2015}.

Recently, De Vos \textit{et al.} \cite{deVos2019DLIR} proposed a Deep Learning Image Registration (DLIR) framework for unsupervised affine and deformable image registration.
This framework consists of a multi-stage CNN architecture for the coarse-to-fine registration considering multiple levels and image resolutions and achieved comparable performance with respect to conventional image registration while being several orders of magnitude faster.
A progressive training method for end-to-end image registration based on a U-Net~\cite{ronneberger2015} was devised in~\cite{eppenhof2019}, which gradually processed from coarse-grained to fine-grained resolution data.
The network was progressively expanded during training by adding higher resolution layers that allowed the network to learn fine-grained deformations from higher-resolution data.

The starting point of the present work was the VoxelMorph framework that was recently introduced for deformable registration of brain Magnetic Resonance Imaging (MRI) images and is considered state-of-the-art~\cite{balakrishnan2019}.
The VoxelMorph framework is fully unsupervised and allows for a clinically feasible real-time solution by registering full 3D volumes in a single-shot.
From a research perspective, the framework is flexible to modifications and extensions of the network architecture.
VoxelMorph formulates the registration as a parameterized function $g_\theta(\cdot,\cdot)$ learned from a collection of volumes in order to estimate the DVF $\phi$.
This parameterization $\theta$ is based on a CNN architecture similar to U-Net~\cite{ronneberger2015} which allows for the combination of low- and high-resolution features, and is estimated by minimizing a loss function using a training set.
The initial VoxelMorph model was evaluated on a dataset of $7829$ T1-weighted brain MRI images acquired from eight different public datasets.
As extensions of this model, Kim \textit{et al.}~\cite{kim2019unsupervised} integrated cycle-consistency~\cite{zhu2017cycle} into VoxelMorph, showing that even image pairs with severe deformations can be registered by improving topology preservation.
In addition, the combination of VoxelMorph with FlowNet~\cite{dosovitskiy2015} for motion correction of respiratory-gated Positron Emission Tomography (PET) scans was proposed in~\cite{li2019}.

\section{Materials and methods}
\label{sec:MatMet}

\subsection{Dataset description}
\label{sec:dataset}

The dataset used in this study comprised consecutive CT images of patients with bone metastases originating from primary breast cancer.
Breast cancer frequently presents with a mixture of lytic and sclerotic bone metastases, where lytic metastases appear similar to areas of low Hounsfield Unit (HU) attenuation in the bones and sclerotic metastases are more densely calcified than normal bone and have higher HU attenation.
Treatment response often causes increasing sclerosis, especially in lytic metastases.
However, increasing sclerosis can also be a sign of disease progression, especially in patients with mixed or purely sclerotic metastases at diagnosis, thus causing a diagnostic dilemma \cite{burns2013}.
Quantitative assessment of bone metastases and the associated changes in attenuation and bone texture over time thus holds the potential to improve treatment response assessment \cite{yankeelov2016,blackledge2014,reischauer2018}.
To enable such assessments, accurate and preferably real-time deformable registration of the consecutive CT images is an important prerequisite.

After informed consent, patients with metastatic breast cancer were recruited into a study designed to characterize the disease at the molecular level, using tissue samples and serial samples of circulating tumor DNA (ctDNA)~\cite{mouliere2018}.
CT imaging of the chest, abdomen, and pelvis was acquired according to clinical request every $3$-$12$ months to assess response to standard-of-care treatment.
A subset of $12$ patients with bone metastases only were selected, resulting in $88$ axial CT images of the abdomen and pelvis.
The CT images were acquired using either of two different clinical CT scanner models---the SOMATOM Emotion 16, the SOMATOM Definition AS(+), and  the SOMATOM Sensation 16---manufactured by Siemens Healthineers (Erlangen, Germany).

On axial images reconstructed with a slice thickness of $2$ mm and a pixel spacing ranging from $0.57$-$0.97$ mm using bone window settings, all vertebral bodies of the thoracic and lumbar spine that were depicted completely were segmented semi-automatically by a board certified radiologist with ten years of experience in clinical imaging, using Microsoft Radiomics (project InnerEye\footnote[2]{\url{https://www.microsoft.com/en-us/research/project/medical-image-analysis/}}, Microsoft, Redmond, WA, USA).
Thus, a series of closely neighboring VOIs was created that spanned the majority of the superior-inferior extent of each scanning volume and was used subsequently to assess the performance of the registration approach.
The total number of VOIs delineated for the analyzed dataset was $805$ (mean VOIs per scan: $9.15$).

\subsection{Dataset preparation and training set construction}
\label{sec:dataPrepConstr}

\subsubsection{Abdominopelvic CT image pre-processing}
\label{sec:dataPrep}

\paragraph{CT table removal}
\label{sec:CT_tableRem}

In a manner similar to that of the commonly used data preparation procedure for brain MR images called ``skull-stripping'' \cite{gambino2011}, we refined our abdominopelvic CT images to facilitate deformable registration.
The CT table could bias the learning process and lead the registration to overfit on the patient table region. Therefore, we developed a fully automatic approach based on region-growing \cite{rundoMBEC2016} to remove the CT table from the CT images, as well as all extra-corporeal components, such as breast prostheses, clothes and metal objects.
Our slice-by-slice approach automatically initialized the growing region, $\mathcal{R}_G$, with a $50 \times 50$-pixel squared seed-region at the center of each slice by assuming that the body was positioned at the center of the CT scanner.

Considering an image $\mathbf{I}$, Eq.~(\ref{eq:RGcriteria}) defines the homogeneity criterion, $\mathsf{P}$, in terms of the mean value of the region $\mu_{R_G}$ \cite{rundoMBEC2016}:

\begin{equation}
    \label{eq:RGcriteria}
    \mathsf{P} =
     \begin{cases}
        \mathsf{True} \text{, if } \mathbf{p}_\mathcal{B} \notin \mathcal{R}_G \wedge |\mathbf{I}(\mathbf{p}_B) - \mu_{\mathcal{R}_G}| < T_G\\
        \mathsf{False} \text{, otherwise}
     \end{cases},
\end{equation}
where $\mathbf{p}_\mathcal{B} \in \mathcal{B}$ denotes a pixel belonging to the candidate list $\mathcal{B}$ of the boundary pixels in the growing region $\mathcal{R}_G$, while $T_G$ is the inclusion threshold.
In particular, during the iterations, the $8$-neighbors of the current pixel $\mathbf{p}_\mathcal{B}$, which do not yet belong to $\mathcal{R}_G$, are included into the candidate list $\mathcal{B}$.
The similarity criterion, $\mathsf{P}$, was based on the absolute difference between the value of the candidate pixels $\mathbf{I}(\mathbf{p})$ and the mean intensity of the pixels included in $\mathcal{R}_G$ (i.e., $\mu_{\mathcal{R}_G} = \sum_{\mathbf{q} \in \mathcal{R}_G} \mathbf{I}(\mathbf{q})/|\mathcal{R}_G|$.
If this difference is lower than $T_G$, the current pixel $\mathbf{p}$ under consideration is added to $\mathcal{R}_G$.
The procedure ends when the list $\mathcal{B}$ is empty.
To account for the variability of the different CT scans, the inclusion threshold, $T_G$, is incrementally increased until $|\mathcal{R}_G|$ reaches a minimum area of $6000$ pixels.
In more details, the input CT pixel values (expressed in HU) are transformed into the range $[0,1]$ (\textit{via} a linear mapping) and the value of $T_G$ varies in $[0.08,0.4]$ at $0.02$ incremental steps at each iteration.
Finally, all automated refinements were carefully verified.

\begin{figure}[!ht]
	\centering
	\subfloat[]{\includegraphics[width=0.45\textwidth]{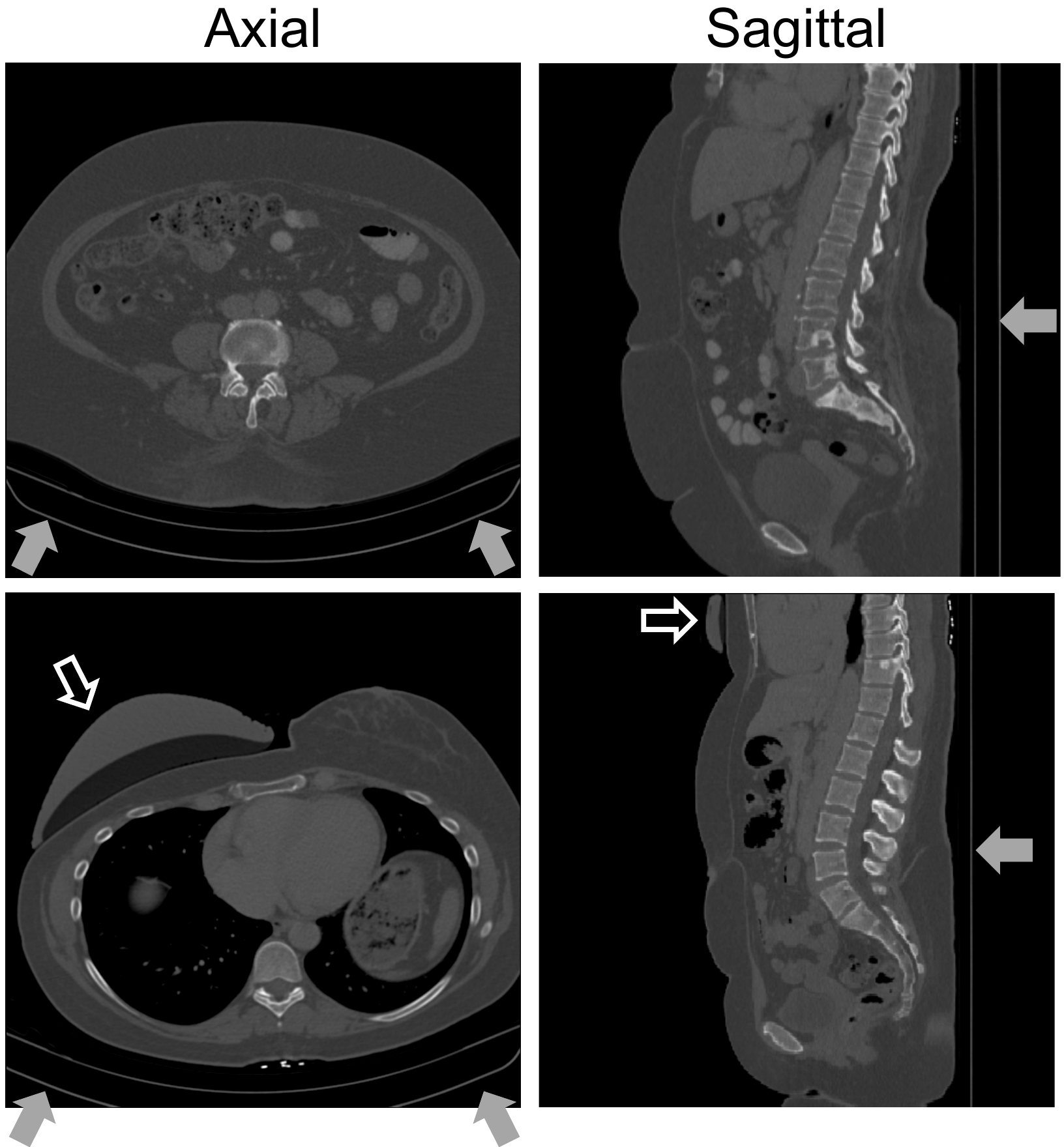}\label{sfig:TableRem_orig}} \quad
	\subfloat[]{\includegraphics[width=0.45\textwidth]{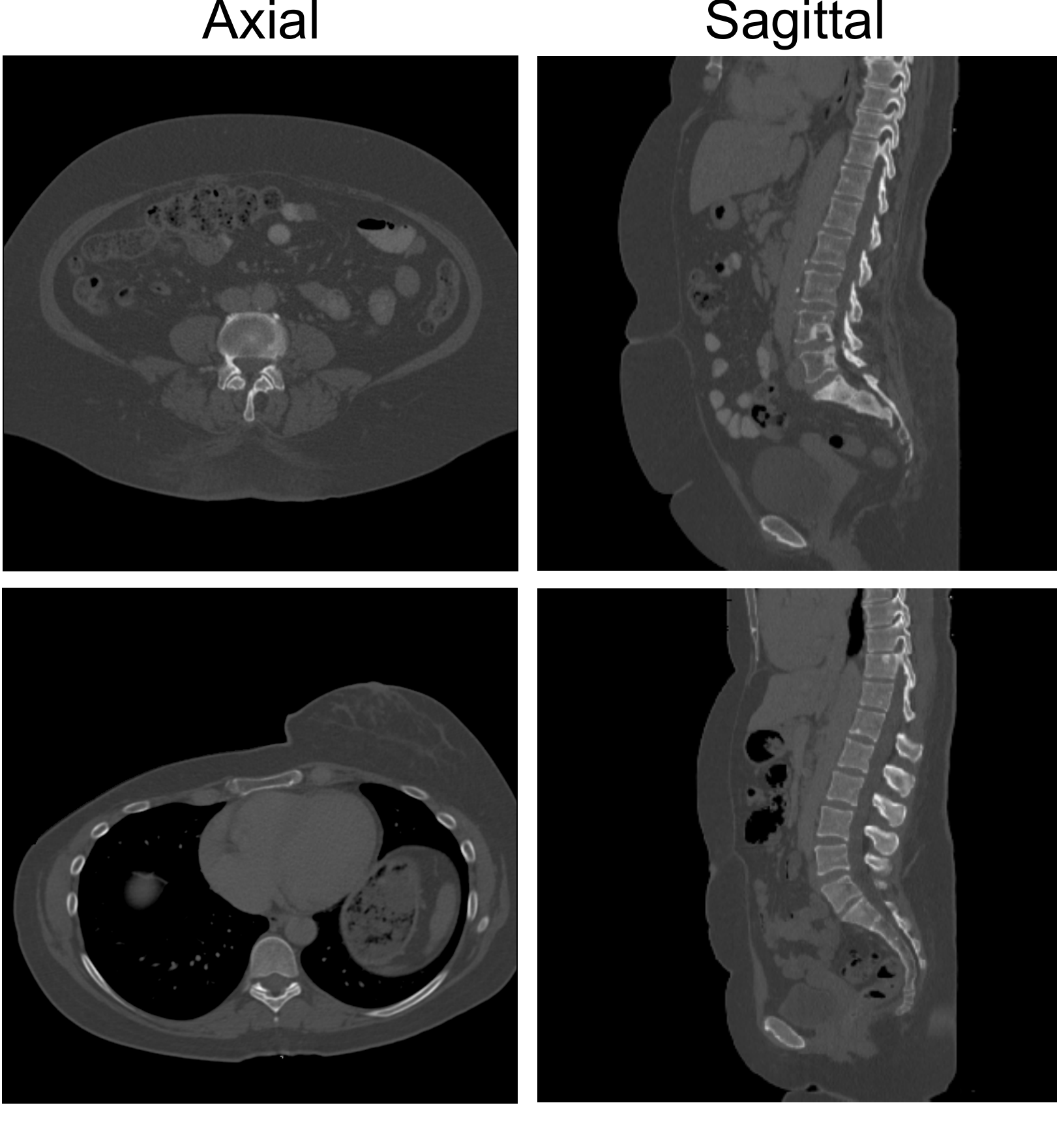}\label{sfig:TableRem_ref}}\\
	\caption{Two example pairs of input axial and sagittal CT slices from the analyzed dataset: (a) original images; (b) refined images where the CT table and other extra-corporeal parts were removed. The CT table and the breast prosthesis are indicated by solid gray and empty white arrows, respectively. Window level and width are set to $400$ and $1800$ HU, respectively, optimized for spine bone visualization.}
	\label{fig:TableRemoval}	
\end{figure}

Fig.~\ref{fig:TableRemoval} shows two examples of CT table removal.
In particular, the sagittal view shows how the CT table was removed along the whole scan (Fig.~\ref{sfig:TableRem_ref}).
In addition, the extra-corporeal parts (i.e., breast prostheses) are discarded in  the second example (bottom row).

\paragraph{CT image pre-processing}
After CT table removal, the following additional data pre-processing steps were performed:
\begin{enumerate}
    \item Affine registration using the NiftyReg toolbox~\cite{modat2010} to account for global rotations and translations, as well as scaling factors in the case of different Fields-of-View (FOVs);
    \item Normalization per scan in $[0,1]$ by means of linear stretching to the $99$th percentile:
    $\tilde{x}_{i} = \frac{x_{i} - x_{\min}}{x_{\max} - x_{\min}}$ for $i \in \{x_{\min}, x_{\min}+1, \ldots, x_{\max}\}$;
    \item Downsampling with a factor of $2$ to $160 \times 160 \times 256$ ($1 \mbox{ mm}^3$ voxels) and cropping to the convex hull (box) enclosing all volumes.
\end{enumerate}

\subsubsection{Generation of simulated DVFs}
\label{sec:DefFieldGen}

It was not possible to directly train a network to register the longitudinal abdominopelvic CT images in our dataset due to the limited amount of available transformation pairs (see Section \ref{sec:dataset}), large inter-patient variations, and the often non-diffeomorphic nature of the transformations, e.g., due to the changes in the appearances of normal structures in consecutive CT images caused by bowel peristalsis or bladder filling.
Therefore, we developed a simulator that generated random synthetic DVFs and transforms abdominopelvic CT images in a manner similar to that of Sokooti \textit{et al.}~\cite{sokooti2017} and Eppenhof \textit{et al.}~\cite{eppenhof2019TMI}.
The resulting deformed CT images can subsequently be used to train or evaluate deep learning based image registration methods.

\begin{figure}[H]
    \centering
    \includegraphics[width=\textwidth]{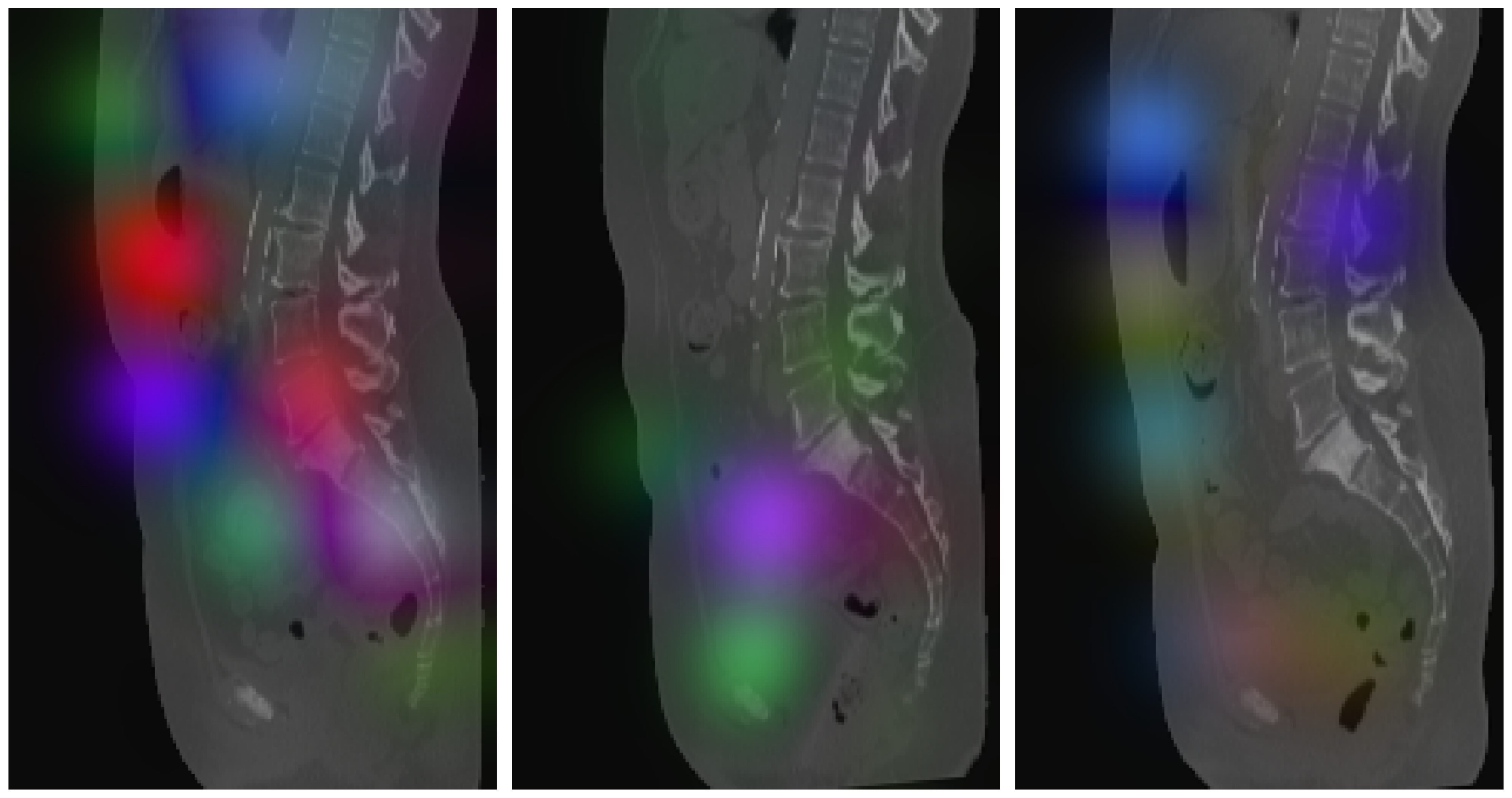}
    \vspace{-1cm}
    \caption{Randomly selected examples of simulated DVFs (same patient; first three time-points). The displacements in the $x$, $y$, and $z$ spatial directions are encoded by the Red, Green, and Blue (RGB) color channels of an RGB image superimposed on the corresponding sagittal CT image \textit{via} alpha blending.}
    \label{fig:DVFex}
\end{figure}

The synthetic DVF generator randomly selects $P$ initialization points, $\mathbf{d}_i$ (with $i=1,2, \ldots, P$), from within the patient volume of a CT image with a minimum distance, $d_P$, between these points.
In the present study, all DVFs were generated using $P=100$ and $d_P=40$.
Each point, $\mathbf{d}_i$, is composed of three random values between $-\delta$ and $\delta$ that correspond to the $x$, $y$, and $z$ components of the displacement vector in that point.
To ensure that the simulated displacement fields were as realistic as possible, we set  $\delta = 6$ to mimic the typical displacements found between the pre-registered images in our abdominopelvic CT dataset.
In addition, we generated a series of DVFs with increasingly large displacements ($\delta = [0,1, \ldots, 25]$) for evaluation purposes (see Section~\ref{sec:largeDispl}).
The resulting vectors were subsequently used to initialize a displacement field, $\phi_s$, with the same dimensions as the original CT image.
To ensure that the DVF moved neighboring voxels into the same direction, the displacement field was smoothed with a Gaussian kernel with a standard deviation of $\sigma_s =0.005$.
Three examples of resulting synthetic DVFs are shown in Fig.~\ref{fig:DVFex}.
Finally, the CT image was transformed using the generated DVF and Gaussian noise with a standard deviation of $\sigma_n = 0.001$, which was added to make the transformed CT image more realistic.
The resulting deformed CT images had a mean Dice Similarity Coefficient (DSC) of $0.725\pm0.059$, which corresponded to the initial differences between the real scan pairs in our longitudinal abdominopelvic CT dataset (see Fig.~\ref{fig:ResRealDefs}). A detailed explanation of DSC can be found in Section~\ref{sec:evalMethod}.

\subsection{Deep learning based deformable image registration}
\label{sec:proposedMethod}

\subsubsection{The VoxelMorph framework}
\label{sec:VoxelMorph}

\begin{figure}[ht]
    \hspace{-2.3cm}
    \includegraphics[width=1.3\textwidth]{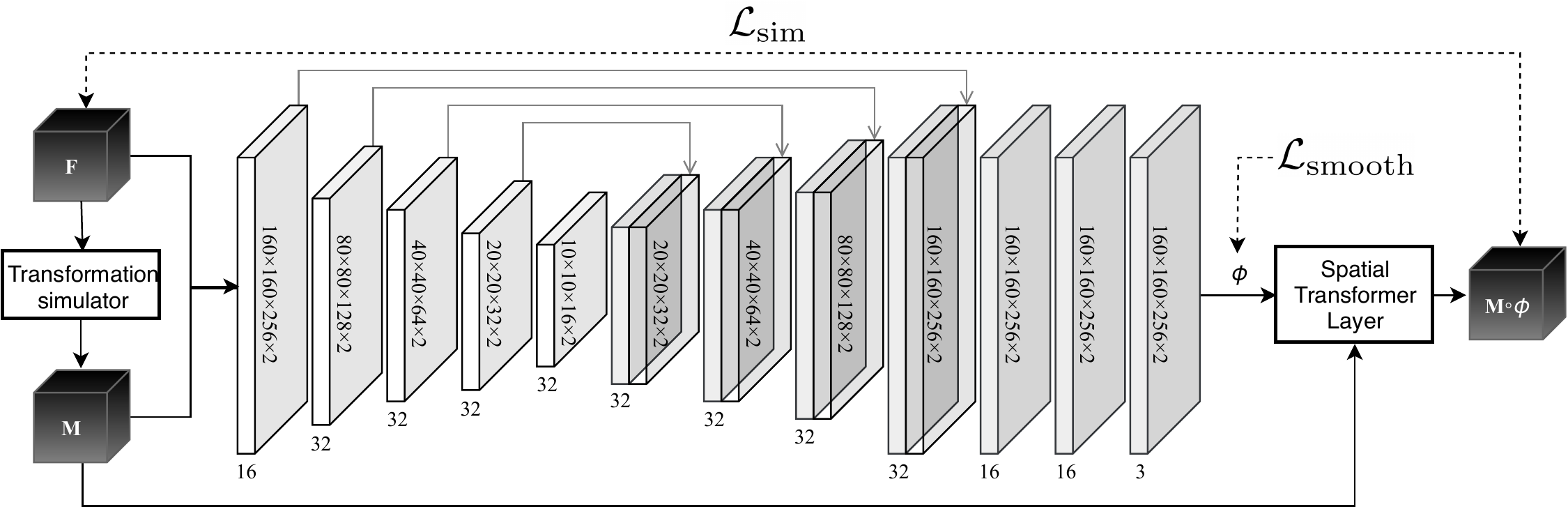}
    \vspace{-1cm}
    \caption{CNN architecture implementing $g_\theta(\mathbf{F},\mathbf{M})$ based on VoxelMorph~\cite{balakrishnan2019}. The spatial resolution of the input 3D volume of each 3D convolutional layer is shown vertically, while the number of feature maps is reported below each layer.
    The black solid lines denote the operations that involve the input fixed $\mathbf{F}$ and moving $\mathbf{M}$ volumes, while the black dashed lines represent the arguments of the loss function components $\mathcal{L}_\text{sim}$ and $\mathcal{L}_\text{smooth}$.}
    \label{fig:NetArchitecture}
\end{figure}

The VoxelMorph model consists of a CNN that takes a fixed and a moving volume as input, followed by an STL that warps the moving volume using the deformation that is yielded by the CNN (Fig.~\ref{fig:NetArchitecture}).
The model can be trained with any differentiable loss function.
Let $\mathbf{F}$ and $\mathbf{M}$ be two image volumes defined over an $N$-dimensional spatial domain, $\Omega \subset \mathbb{R}^N$.
We consider CT images, thus $N=3$ in our study.
More specifically, $\mathbf{F}$ and $\mathbf{M}$ were the fixed and moving images, respectively.

Let $\phi$ be a transformation operator defined by a DVF $\mathbf{u}$ that denotes the offset vector from $\mathbf{F}$ to $\mathbf{M}$ for each voxel: $\phi = \text{Id} + \mathbf{u}$, where $\text{Id}$ is the identity transform.
We used the following unsupervised loss function:
\begin{equation}
    \label{eq:UnsupLoss}
    \mathcal{L}(\mathbf{F},\mathbf{M};\phi) = \mathcal{L}_\text{sim}(\mathbf{F},\mathbf{M} \circ \phi) + \lambda \mathcal{L}_\text{smooth}(\phi),
\end{equation}
where $\mathcal{L}_\text{sim}$ aims to minimize differences in appearance and $\mathcal{L}_\text{smooth}$ penalizes the local spatial variations in $\phi$, acting as a regularizer weighted by the parameter $\lambda$.
The employed $\mathcal{L}_\text{sim}$ is the local cross-correlation between $\mathbf{F}$ and $\mathbf{M} \circ \phi$, which is more robust to intensity variations found across scans and datasets~\cite{avants2008}.
Let $\hat{\mathbf{F}}(\mathbf{p})$ and $[ \hat{\mathbf{M}} \circ \phi ]({\mathbf{p}})$ denote local mean intensity images: {$\hat{\mathbf{F}}({\mathbf{p}}) = \frac{1}{\omega^3}\sum_{\mathbf{p}_i \in \mathcal{N}(\mathbf{p})} \mathbf{F} (\mathbf{p}_i)$}, where $\mathbf{p}_i$ iterates over a local neighborhood, $\mathcal{N}(\mathbf{p})$, defining an $\omega^3$ volume centered on $\mathbf{p}$, with $\omega=9$ in our experiments.
The local cross-correlation of $\mathbf{F}$ and $[\mathbf{M} \circ \phi]$ is defined as:
\begin{align}
    & \text{NCC}(\mathbf{F}, \mathbf{M} \circ \phi)= \nonumber \\
    & \sum\limits_{\mathbf{p} \in \Omega} \frac{\left(\sum\limits_{\mathbf{p}_i \in \mathcal{N}(\mathbf{p})} (\mathbf{F} ({\mathbf{p}_i}) - \hat{\mathbf{F}} ({\mathbf{p}}))([\mathbf{M} \circ \phi] ({\mathbf{p}_i}) - [\hat{\mathbf{M}} \circ \phi] ({\mathbf{F}}))\right)^2}
    {\left(\sum\limits_{\mathbf{p}_i \in \mathcal{N}(\mathbf{p})} (\mathbf{F} ({\mathbf{p}_i}) - \hat{\mathbf{F}} ({\mathbf{p}}))^2 \right)\left(\sum\limits_{\mathbf{p}_i \in \mathcal{N}(\mathbf{p})} ([\mathbf{M} \circ \phi] ({\mathbf{p}_i}) - [\hat{\mathbf{M}} \circ \phi] ({\mathbf{p}}))^2\right)}. \label{eq:NCC}
\end{align}

\noindent A higher NCC indicates a better alignment, yielding the loss function:
\begin{equation}
    \label{eq:Lsim}
    \mathcal{L}_\text{sim}(\mathbf{F},\mathbf{M};\phi) = - \text{NCC}(\mathbf{F}, \mathbf{M} \circ \phi).
\end{equation}

Minimizing $\mathcal{L}_\text{sim}$ encourages $\mathbf{M} \circ \phi$ to approximate~$\mathbf{F}$, but might yield a non-smooth $\phi$ that is not physically realistic.
Thus, a smoother displacement field $\phi$ is achieved by using a diffusion regularization term on the spatial gradients of displacement~$\mathbf{u}$:

\begin{equation}
    \label{eq:Lsmooth}
    \mathcal{L}_\text{smooth}(\phi) = \sum_{\mathbf{p} \in \Omega} ||\nabla {\mathbf{u}(\mathbf{p})}||^2,
\end{equation}
and approximate spatial gradients \textit{via} the differences among neighboring voxels. 

Fig.~\ref{fig:NetArchitecture} depicts the CNN used in VoxelMorph, which takes a single input formed by concatenating  $\mathbf{F}$ and  $\mathbf{M}$ into a two-channel 3D image.
Taking inspiration from U-Net~\cite{ronneberger2015}, the decoder uses several $32$-filter convolutions, each followed by an upsampling layer, to bring the volume back to full-resolution. The gray lines denote the skip connections, which concatenate coarse-grained and fine-grained features.
The full-resolution volume is successively refined \textit{via} several convolutions and the estimated deformation field, $\phi$, is applied to the moving image, $\mathbf{M}$, \textit{via} the STL~\cite{jaderberg2015}.
In our experiments, the input was $160 \times 160 \times 256 \times 2$ in size.
3D convolutions were applied in both the encoder and decoder paths using a kernel size of $3$, and a stride of $2$.
Each convolution was followed by a Leaky Rectified Linear Unit (ReLU) layer with parameter $\alpha$.
The convolutional layers captured hierarchical features of the input image pair, used to estimate $\phi$.
In the encoder, strided convolutions were exploited to halve the spatial dimensions at each layer.
Thus, the successive layers of the encoder operated over coarser representations of the input, similar to the image pyramid used in hierarchical image registration approaches.

\subsubsection{Parameter settings and implementation details}
\label{sec:Impl}
In the present study, the optimized hyperparameter settings suggested by Balakrishnan \textit{et al.}~\cite{balakrishnan2019} served as a starting point.
We investigated the effect of the LeakyReLU $\alpha$ parameter on the stability of the training process and found that an $\alpha$ of $0.5$ was optimal for registering abdominopelvic CT images.
In all experiments, the regularization parameter, $\lambda$, was set to $1.0$.
One training epoch consisted of $100$ steps and took approximately five minutes.
The models described in Section~\ref{sec:tabRem_Res} were trained until convergence ($1000$ epochs) using a learning rate of $10 \times 10^{-4}$, whereas the models described in Section~\ref{sec:regAcc_Res} were trained using the early stopping monitoring function implemented in the Python programming language using Keras (with a TensorFlow backend) based on $50$ validation steps and a patience of $20$ epochs.
Training was parallelized on four Nvidia GeForce GPX 1080 Ti (Nvidia Corporation, Santa Clara, CA, USA) GPUs (batch size = $4$) and evaluation of the trained networks was performed using an Nvidia GeForce GPX 1070 Ti GPU. 

\subsubsection{Incremental training strategy}
\label{sec:incrTraining}

\begin{figure}[ht]
    \hspace{-2.3cm}
    \includegraphics[width=1.3\textwidth]{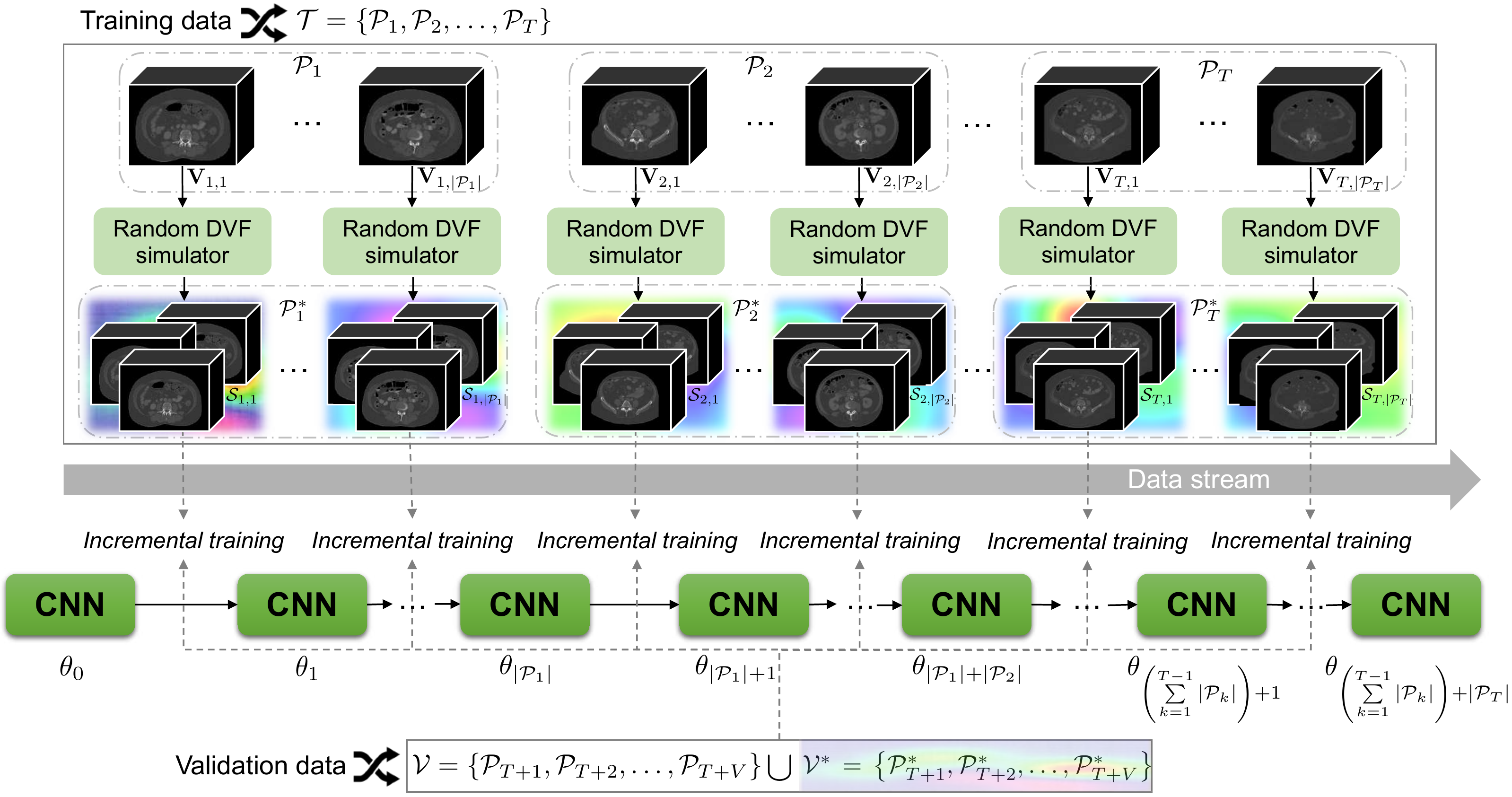}
    \vspace{-1cm}
    \caption{Workflow of the proposed incremental training strategy: $\mathcal{T}$ and $\mathcal{V}$ represent the training and validation sets, respectively. The parameters $\theta$, employed in the parameterized registration functions $g_{\theta}(\cdot,\cdot)$, are incrementally learned for each deformed volume included in the training set $\mathcal{T}$ and tested on the unseen volumes of the validation set $\mathcal{V}$. All deformed volumes in $\mathcal{T}$ and $\mathcal{V}$ are synthesized using a random DVF simulator. The notation $\mathbf{V}_{i,j}$ denotes the $j$-th 3D volume for a patient, $\mathcal{P}_i$ (with $i \in \{1, 2, \ldots D\}$ and $D = T+V$).}
    \label{fig:flowchart}
\end{figure}

The VoxelMorph network did not converge when it was na\"{i}vely trained on the limited number of abdominopelvic CT scans in the available dataset $\mathcal{D}$ (only $76 \times 2 = 152$ possible intra-patient combinations). 
To overcome this limitation, we developed a novel approach to enforce learning based on simulated deformations (see Section~\ref{sec:DefFieldGen}) and incremental learning.
The term incremental learning refers to learning from a constantly arriving datastream, which is an important research topic in computer vision and pattern recognition, such as autonomous robotics or driving \cite{chefrour2019}.
In our incremental training strategy (Fig.~\ref{fig:flowchart}), deformed CT images are sequentially presented to the network in chronological mini-batches per patient.

Let $\mathcal{D} = \left\{ \mathcal{P}_1, \mathcal{P}_2, \ldots, \mathcal{P}_D \right\}$ contain all abdominopelvic CT images for each patient $\mathcal{P}_i = \left\{ \mathbf{V}_{i,1}, \mathbf{V}_{i,2}, \ldots, \mathbf{V}_{i,|\mathcal{P}_i|} \right\}$, where $i=1,2, \ldots, D$ and $|\mathcal{P}_i| $ denotes the patient index and the corresponding number of CT volumes, respectively.
The whole dataset, $\mathcal{D}$, was split into two disjoint training, $\mathcal{T} = \left\{ \mathcal{P}_1, \mathcal{P}_2, \ldots, \mathcal{P}_T \right\}$, and validation, $\mathcal{V} = \left\{ \mathcal{P}_{T+1}, \mathcal{P}_{T+2}, \ldots, \mathcal{P}_{T+V}, \right\}$ sets with $T+V = D$.
In our case, $D=12$ with $T=9$ and $V=3$.
Each volume, $\mathbf{V}_{i,j}$ (with $j=1,2,\ldots,|\mathcal{P}_i|$), was subsequently deformed using $K$ randomly generated DVFs, $\phi_k$ (see Section \ref{sec:DefFieldGen}), resulting in $\mathcal{S}_{i,j}=\left\{ \mathbf{V}_{i,j}^{(k)} \right\}_{k=1,\ldots,K}$ deformed volumes for the $i$-th patient, with $i=1,2,\ldots,D$.

The set $\mathcal{T}^* = \left\{ \mathcal{P}_1^*, \mathcal{P}_2^*, \ldots, \mathcal{P}_T^* \right\}$, with $\mathcal{P}_i^* = \left\{\mathcal{S}_{i,1}, \mathcal{S}_{i,2}, \ldots. \mathcal{S}_{i,|\mathcal{P}_i|} \right\}$, was used to incrementally train the network such that in each training iteration the network was trained on a mini-batch containing all deformed volumes, $\mathcal{S}_{i,j}$.
The deformed volumes in the set $\mathcal{V}^* = \left\{ \mathcal{P}_{T+1}^*, \mathcal{P}_{T+2}^*, \ldots, \mathcal{P}_{T+V}^* \right\}$ were randomly divided into two equal, independent parts.
One part was kept aside for evaluation, and the other part was used to monitor the training process to avoid concept drift (i.e., changes in the data distribution) between the mini-batches over time.
After each training iteration, the network weights that resulted in the best performance on this second part of $\mathcal{V}^*$ were reloaded to initiate the next iteration. 
If the network did not converge during a certain iteration, the network weights of the previous iteration were reloaded, thereby ensuring that the overall training process could continue and remain stable.
To reduce forgetting, the learning rate was decreased linearly from from $10^{-4}$ (first iteration) to $10^{-6}$ (last iteration) \cite{goyal2017}.

The incremental training strategy was evaluated using a $4$-fold cross-validation scheme in which all patients in dataset $\mathcal{D}$ were randomly shuffled while the order of the distinct time-points was preserved in order to account for the longitudinal nature of our dataset.
Since $D=12$, $\sum\limits_{i=1,2,\ldots,D} |\mathcal{P}_i| = 88$, and $K=30$, a total of $2640$ deformed volumes, $\mathcal{D}^*$, were generated in this study, of which $2014$ were used for training, $323$ for monitoring the training process, and $323$ for evaluation in each cross-validation round.
Cross-validation allows for a better estimation of the generalization ability of our training strategy compared to a hold-out method in which the dataset is partitioned into only one training and evaluation set.

\subsection{Evaluation methodology}
\label{sec:evalMethod}

This section describes the evaluation metrics used to quantify the registration performance of the incrementally trained VoxelMorph framework and the NiftyReg toolbox \cite{modat2010} that served as a benchmark in this study.

\subsubsection{NiftyReg}
\label{sec:NiftyReg}
All deformed abdominopelvic CT images were also registered using the Fast Free-Form Deformation (F3D) algorithm for non-rigid registration in the NiftyReg toolbox (version 1.5.58) \cite{modat2010}.
All options were set to default: the image similarity metric used was Normalized Mutual Information (NMI) with $64$ bins and the optimization was performed using a three-level multi-resolution strategy with a maximum number of iterations in the final level of $150$.
Note that the F3D algorithm in the NiftyReg toolbox does not support GPU acceleration, in contrast to the Block Matching algorithm for global (affine) registration in the NiftyReg toolbox that was used to pre-align the CT images in this study (see Section \ref{sec:dataPrep}).

\subsubsection{Evaluation metrics}
\label{sec:metrics}
To quantify image registration performance, we relied on highly accurate delineations performed by a board-certified radiologist.
The rationale for considering the VOIs that covered the vertebral bodies of the spine to determine registration performance was that they spanned the majority of the scanning volume in the superior-inferior direction and were of clinical relevance because of the underlying study on bone metastases.

As an evaluation metric, we used the DSC, which is often used in medical image registration \cite{balakrishnan2019}.
DSC values were calculated using the gold standard regions delineated on the fixed scans ($\mathcal{R}_\mathbf{F}$) and the corresponding transformed regions on the moving scans ($\mathcal{R}_\mathbf{M}$) after application of the estimated DVF $\phi$: $\mathcal{R}_\mathbf{D} = \mathcal{R}_\mathbf{M} \circ \phi$ (Eq.~(\ref{eq:DSC})):
\begin{equation}
    \label{eq:DSC}
    \text{DSC} = \frac{2 \cdot |\mathcal{R}_\mathbf{D} \cap \mathcal{R}_\mathbf{F}|}{|\mathcal{R}_\mathbf{D}| + |\mathcal{R}_\mathbf{F}|}.
\end{equation}
Since DSC is an overlap-based metric, the higher the value, the better the segmentation results.

For completeness, we also calculated the Structural Similarity Index (SSIM).
This metric is commonly used to quantify image quality perceived as variations in structural information \cite{wang2004}.
Let $\mathbf{X}$ and $\mathbf{Y}$ be two images (in our case, $\mathbf{F}$ was compared with either $\mathbf{M}$ or $\mathbf{D}$ for the evaluation), and SSIM combines three relatively independent terms:
\begin{itemize}
	\item the luminance comparison  $l(\mathbf{X}, \mathbf{Y}) = \frac{2 \mu_{\mathbf{X}} \mu_{\mathbf{Y}} + \kappa_1}{\mu_{\mathbf{X}}^2 + \mu_{\mathbf{Y}}^2 + \kappa_1}$;
	\item the contrast comparison $c(\mathbf{X}, \mathbf{Y}) = \frac{2 \sigma_{\mathbf{X}} \sigma_{\mathbf{Y}} + \kappa_2}{\sigma_{\mathbf{X}}^2 + \sigma_{\mathbf{Y}}^2 + \kappa_2}$;
	\item the structural comparison $s(\mathbf{X}, \mathbf{Y}) = \frac{\sigma_{\mathbf{X} \mathbf{Y}} + \kappa_3}{\sigma_{\mathbf{X}} \sigma_{\mathbf{Y}} + \kappa_3}$;
\end{itemize}
where $\mu_{\mathbf{X}}$, $\mu_{\mathbf{Y}}$, $\sigma_{\mathbf{X}}$, $\sigma_{\mathbf{Y}}$, and $\sigma_{\mathbf{X}\mathbf{Y}}$ are the local means, standard deviations, and cross-covariance for the images $\mathbf{X}$ and $\mathbf{Y}$, while $\kappa_1, \kappa_2, \kappa_3 \in \mathbb{R}^+$ are regularization constants for luminance, contrast, and structural terms, respectively, exploited to avoid instability in the case of image regions characterized by local mean or standard deviation close to zero.
Typically, small non-zero values are employed for these constants; according to \cite{wang2004}, an appropriate setting is $\kappa_1 = (0.01 \cdot L)^2$, $\kappa_2 = (0.03 \cdot L)^2$, $\kappa_3 = \kappa_2/2$, where $L$ is the dynamic range of the pixel values in $\mathbf{F}$.
SSIM is then computed by combining the components described above:
\begin{equation}
\text{SSIM} = l(\mathbf{X}, \mathbf{Y})^\alpha \cdot c(\mathbf{X}, \mathbf{Y})^\beta \cdot s(\mathbf{X}, \mathbf{Y})^\gamma,
\end{equation}
where $\alpha$, $\beta$, $\gamma > 0$ are weighting exponents.
As reported in \cite{wang2004}, if $\alpha = \beta = \gamma = 1$ and $\kappa_3 = \kappa_2/2$, the SSIM becomes:
\begin{equation}
\text{SSIM} = \frac{\left( 2 \mu_{\mathbf{X}} \mu_{\mathbf{Y}}  + \kappa_1 \right) \left( 2 \sigma_{\mathbf{X} \mathbf{Y}} + \kappa_2 \right) } {\left( \mu_{\mathbf{X}}^2 + \mu_{\mathbf{Y}}^2  + \kappa_1 \right) \left( \sigma_{\mathbf{X}}^2 + \sigma_{\mathbf{Y}}^2  + \kappa_2 \right)}.
\end{equation}
Note that the higher the SSIM value, the higher the structural similarity, implying that the co-registered image, $\mathbf{D}$, and the original image $\mathbf{F}$ are quantitatively similar.

\section{Experimental results}
\label{sec:results}

\begin{figure}[!ht]
	\centering
	\includegraphics[width=0.8\textwidth]{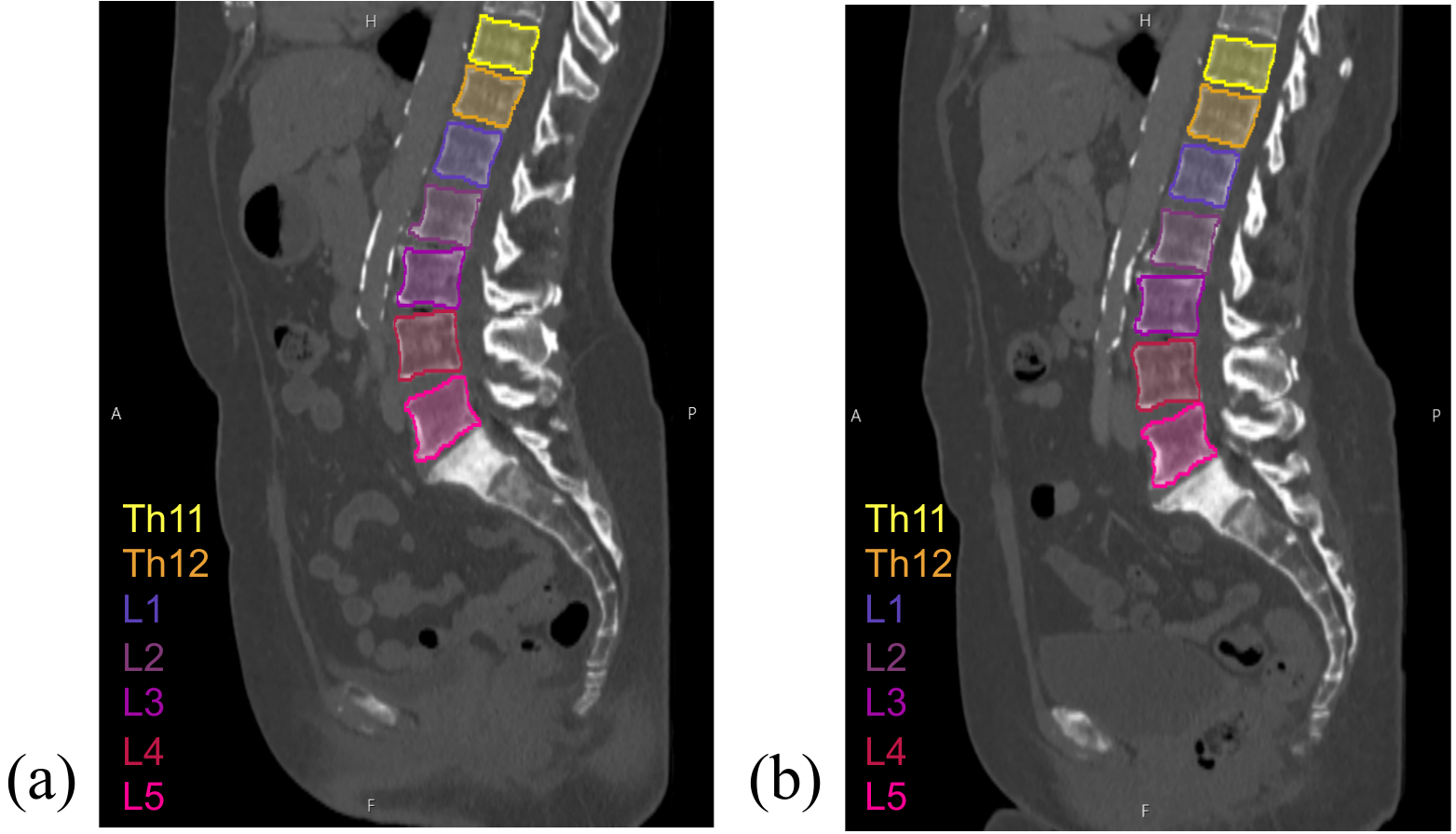}
	\vspace{-0.4cm}
	\caption{Sagittal view of two CT images of the same patient: (a) baseline; (b) second time-point. The vertebrae VOIs are displayed using different colors (legend is shown at the bottom-left). Window level and width are set to $400$ and $1800$ HU, respectively.}
	\label{fig:VOIsagittal}	
\end{figure}

\begin{figure}[!ht]
    \hspace{-2.3cm}
    \includegraphics[width=1.3\textwidth]{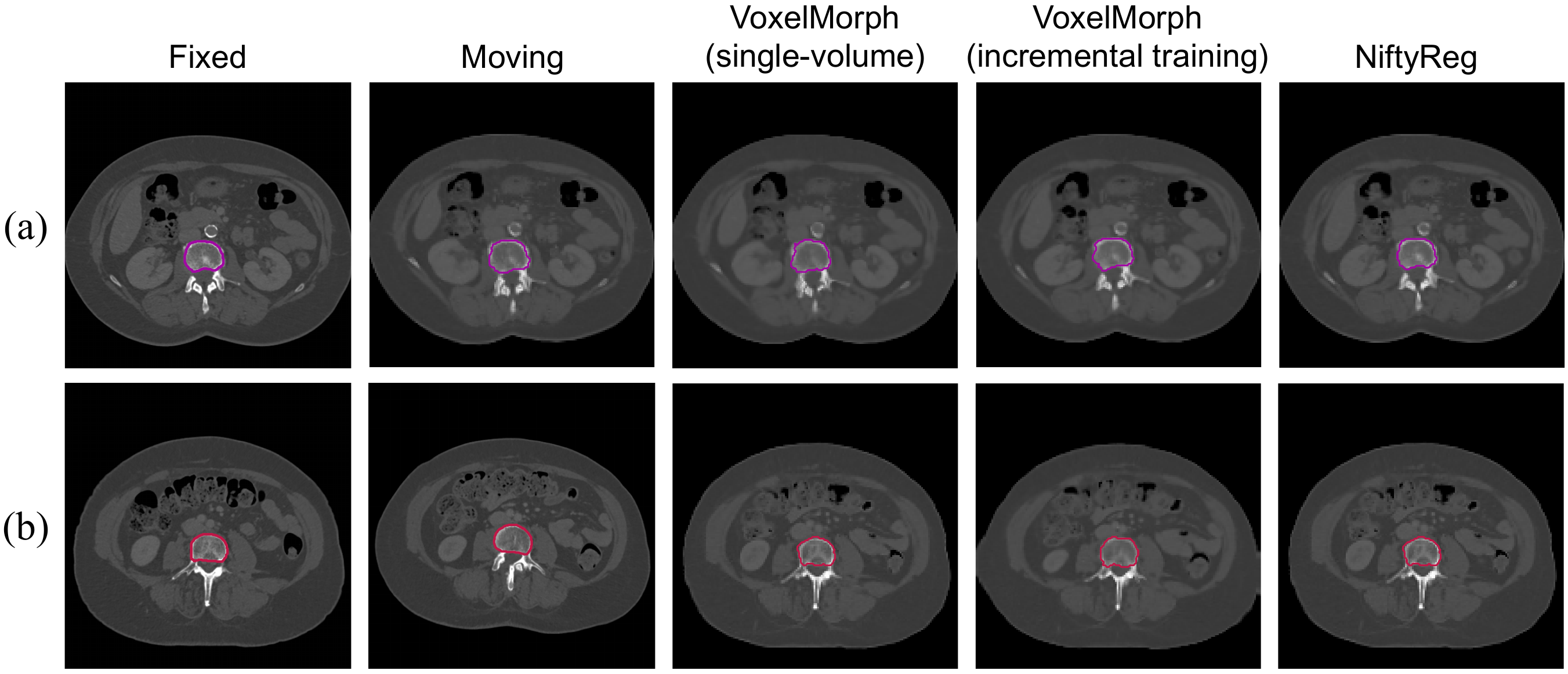}
    \vspace{-1cm}
    \caption{Registration results for the images shown in Fig.~\ref{fig:VOIsagittal} for all the investigated methods: (a) the fixed image is the volume in Fig.~\ref{fig:VOIsagittal}a and the moving image is a simulated deformation of the same volume used during our incremental training procedure; (b) the fixed and the moving images are the volumes in Figs. \ref{fig:VOIsagittal}a and  \ref{fig:VOIsagittal}b, respectively. Example slices with the VOI contours for the L3 and L5 vertebrae are shown in (a) and (b), respectively. Window level and width were set to $400$ and $1800$ HU, respectively.}
    \label{fig:registrationResults}
\end{figure}

Fig.~\ref{fig:VOIsagittal} shows a typical example of two CT images (baseline and second time-point) and VOIs from the same patient from the abdominopelvic CT dataset $\mathcal{D}$. 
Fig.~\ref{fig:registrationResults}a shows an example of deformable registrations achieved using VoxelMorph and NiftyReg in which the moving image was a simulated deformed image (see Section~\ref{sec:DefFieldGen}).
Similarly, Fig.~\ref{fig:registrationResults}b shows an example of a real registration pair from the longitudinal abdominopelvic CT dataset in which the fixed image was the first time-point (Fig.~\ref{fig:VOIsagittal}a) and the moving image was the second time-point (Fig.~\ref{fig:VOIsagittal}b).
Interestingly, the improvement achieved by the proposed incremental training procedure with respect to single-volume training can be appreciated in the VoxelMorph registrations in both Figs.~\ref{fig:registrationResults}a and b.

\subsection{Impact of CT table removal}
\label{sec:tabRem_Res}

\begin{figure}[h]
    \centering
    \includegraphics[width=\textwidth]{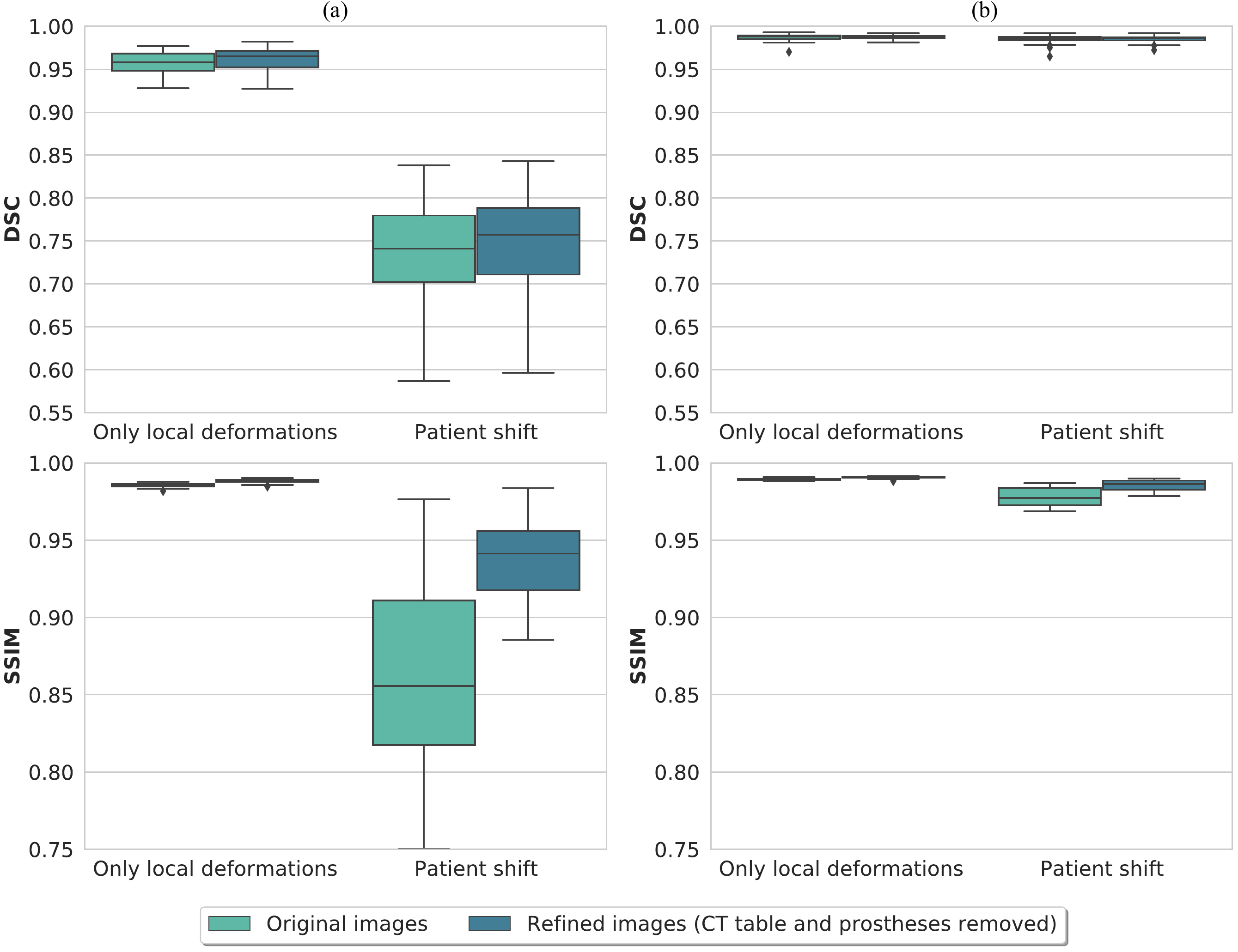}
    \vspace{-1cm}
    \caption{DSC and SSIM of original and refined CT images registered using: (a) VoxelMorph and (b) NiftyReg.}
    \label{fig:CTtableVoxNifty}
\end{figure}

To assess the impact of the CT table removal procedure described in Section~\ref{sec:CT_tableRem} on the image registration performance, $250$ DVFs with a maximum displacement of $5$ mm were randomly simulated such that the initialization points were sampled only from within the patient volume, i.e., the CT table was not deformed. These DVFs were used to deform an original CT scan ($\mathbf{V}_{9,1}$ from $\mathcal{P}_9$) and corresponding refined CT scan (CT table, clothing, and prosthesis removed).
An additional test dataset was created by deforming the original and refined CT scan using both local deformations and a random global translation in the $x$, $y$, and $z$ directions between $-2$ mm and $2$ mm to simulate a small patient shift with respect to the CT table.
Two instances of the VoxelMorph framework were trained on the original and refined datasets, respectively, and tested using $50$ held-out deformed CT images without and with additional global patient shift (Fig.~\ref{fig:CTtableVoxNifty}a).
As a benchmark, all original and refined testing CT images were also registered using NiftyReg (Fig.~\ref{fig:CTtableVoxNifty}b).
Statistical analysis was performed using a paired Wilcoxon signed-rank test, with the null hypothesis that the samples came from continuous distributions with equal medians.
In all tests, a significance level of $0.05$ was set~\cite{wilcoxon1992}.

Fig.~\ref{fig:CTtableVoxNifty}a shows that the VoxelMorph framework achieved significantly higher DSC values when registering refined CT images compared to original CT images for both local deformations ($p<0.005$) and global patient shifts ($p<0.0005$).
Similarly, the SSIM of the refined images registered using the VoxelMorph framework was higher for both local deformations and global patient shifts (both $p<0.0005$).
No difference between original and refined CT images was observed in the DSC values of registrations performed using NiftyReg (Figure~\ref{fig:CTtableVoxNifty}b), although the SSIM of the refined images registered using NiftyReg showed significant improvements over the original images (both $p<0.0005$).

\subsubsection{Computational performance}
\label{sec:compPerf}
Table~\ref{table:computPerform} shows the computational times required to register one image pair using the VoxelMorph framework and NiftyReg.
The CT table removal procedure resulted in slightly shorter registration times when using the NiftyReg toolbox (on average $105$ s) compared to registering original images with and without a patient shift (on average $106$ s and $109$ s, respectively).

\begin{table}[h]
\centering
\tiny
\caption{Computational performance of the deformable registration methods in terms of processing times (mean $\pm$ standard deviation).}
\begin{tabular}{llll}
\hline \hline
\textbf{Method}       &     \textbf{Configuration}         & \textbf{Processing time} [s] \\
\hline
VoxelMorph & all registrations                     & $0.33 \pm 0.015$         \\
\hline
NiftyReg    &   Local deformations                    & $109 \pm 12$        \\
& (original CT scans) & \\
NiftyReg   & Local deformations  & $105 \pm 14$         \\
& (refined CT scans) & \\
NiftyReg          & Local deformations + patient shift                  & $106 \pm 12$         \\
& (original CT scans)  & \\
NiftyReg           & Local deformations + patient shift                     & $105 \pm 5$         \\
& (refined CT scans) &
       \\
\hline \hline
\end{tabular}
\label{table:computPerform}
\end{table}

\subsection{Quantitative evaluation of the incremental training strategy}
\label{sec:regAcc_Res}

\begin{figure}[H]
    \centering
    \includegraphics[width=0.95\textwidth]{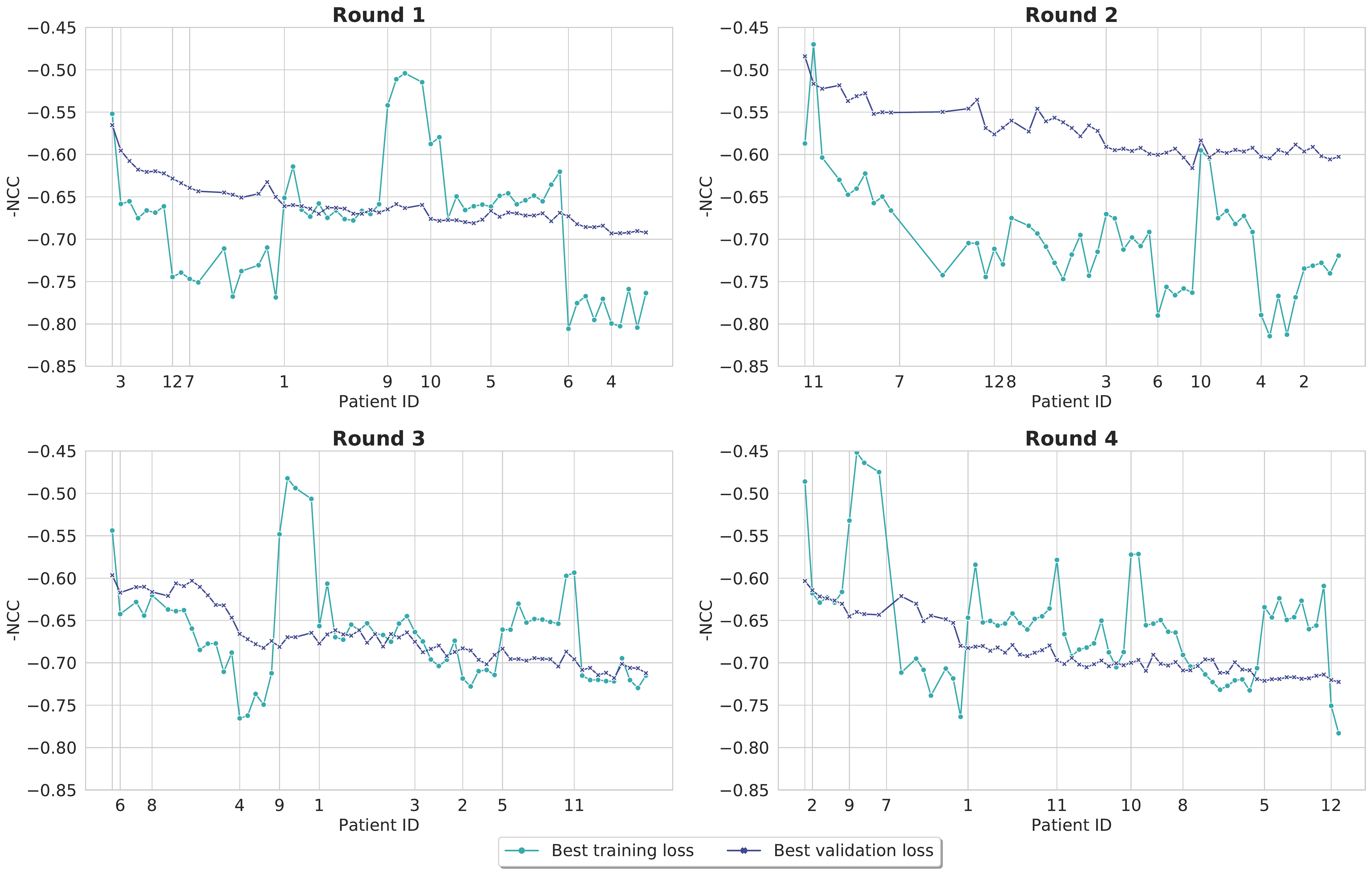}
    \vspace{-0.6cm}
    \caption{Best training and validation errors achieved in each training iteration. The patient IDs represent the randomly initialized order in which the simulated deformed volumes $\mathcal{S}_{i,j}=\left\{ \mathbf{V}_{i,j}^{(k)} \right\}_{k=1,\ldots,K}$ (for the $j$-th scan from the $i$-th patient, with $j=1,2,\ldots,|\mathcal{P}_i|$ and $i=1,2,\ldots,D$) were used during incremental training in each round.}
    \label{fig:IncrTrain}
\end{figure}

In the proposed incremental training strategy, a network was trained on all deformed volumes included in a mini-batch $\mathcal{S}_{i,j}$ until its performance on the validation set $\mathcal{V}^*$ no longer improved, after which the best performing network weights were reloaded to initiate the next training iteration.
Fig.~\ref{fig:IncrTrain} shows the resulting best training and validation errors achieved during each training iteration of the different cross-validation rounds.
Although the training errors sometimes varied greatly between iterations, the network performance on the validation set, $\mathcal{V}^*$, gradually improved during incremental training.

\begin{figure}[!ht]
    \centering
    \includegraphics[width=\textwidth]{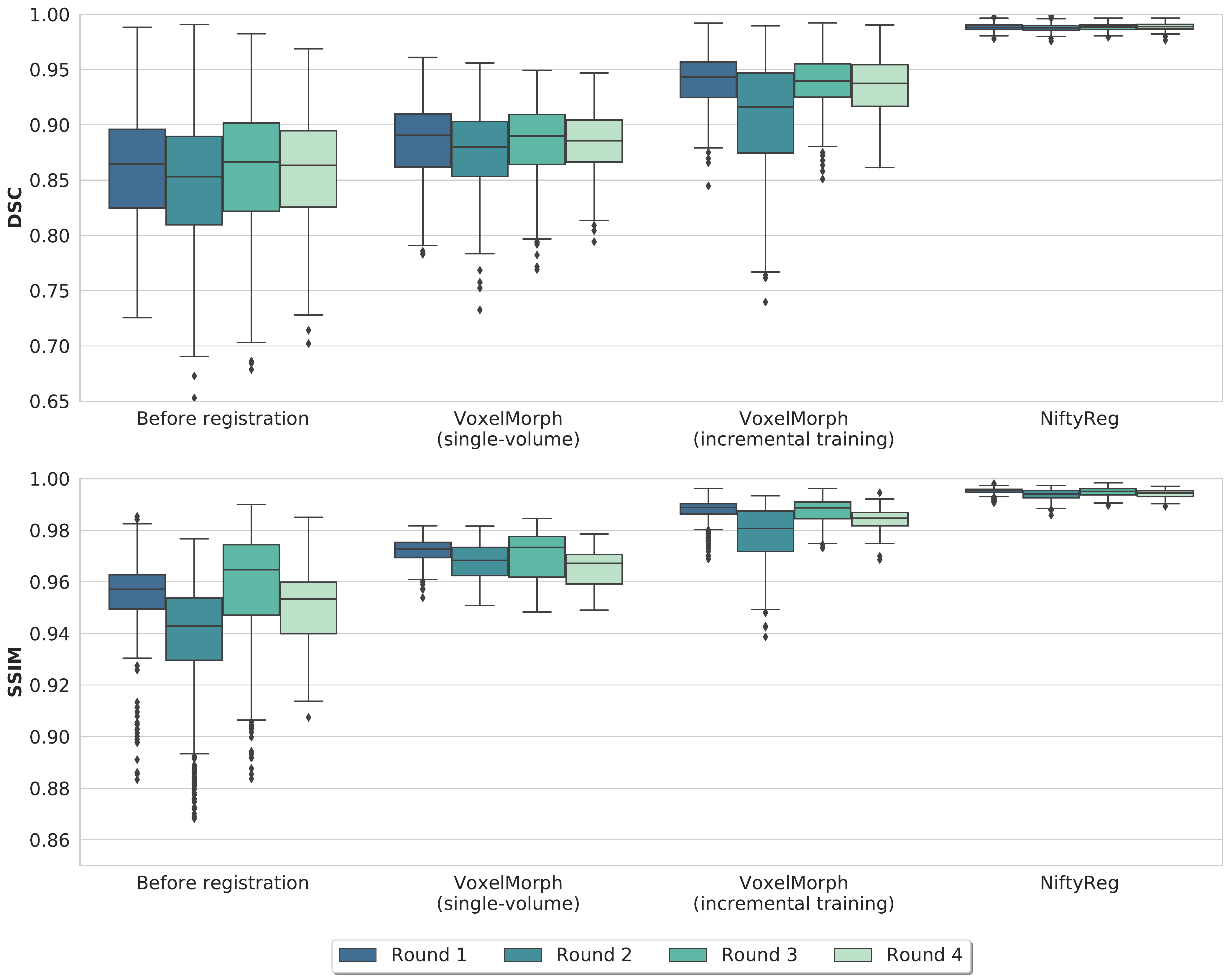}
    \vspace{-1cm}
    \caption{Registration performance on simulated deformations.}
    \label{fig:ResSimDefs}
\end{figure}

\begin{figure}[!h]
    \centering
    \includegraphics[width=\textwidth]{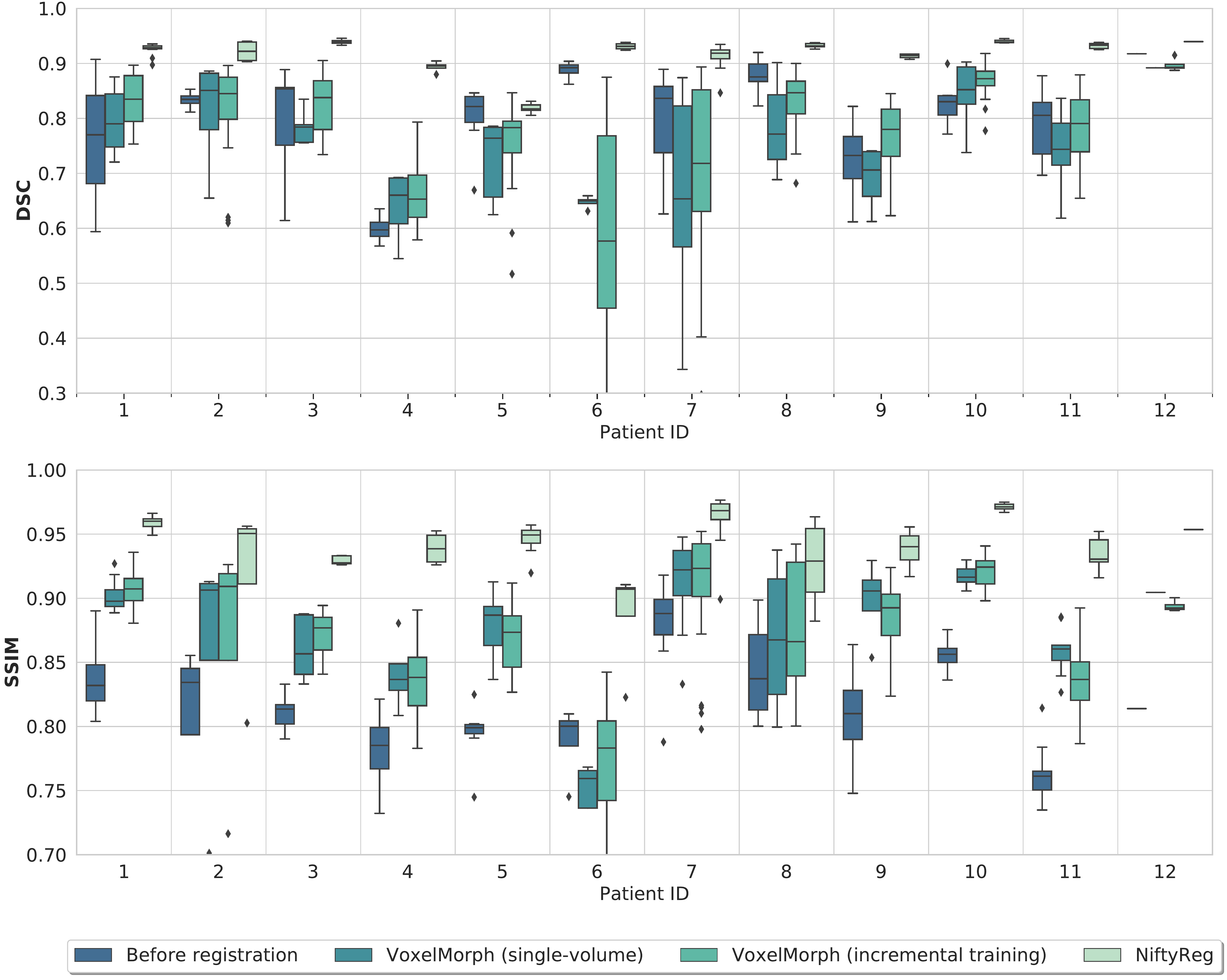}
    \vspace{-1cm}
    \caption{Registration performance on real longitudinal CT images per patient. The incremental training strategy combined all cross-validation rounds.}
    \label{fig:ResRealDefs}
\end{figure}

Another interesting phenomenon that can be observed in Fig.~\ref{fig:IncrTrain} is that the best training errors achievable when training on a specific mini-batch tended to differ between patients.
For example, training errors increased when training on simulated deformed scans of patients $\mathcal{P}_5$ or $\mathcal{P}_9$.
Since both of these patients were, by chance, included in the validation and test sets of round $2$, this also explains why the validation errors in round $2$ were generally higher (Fig.~\ref{fig:IncrTrain}) and the registration performance was lower (see Figs.~\ref{fig:ResSimDefs} and \ref{fig:ResRealDefs}).

\subsubsection{Deformable registration performance}
\label{sec:RegPerf}
Since the VoxelMorph network did not converge when training on either the whole dataset $\mathcal{D}$ or on the simulated dataset $\mathcal{T^*}$, the effectiveness of the proposed incremental training strategy in learning features from multiple patients was compared to training a network on $1000$ simulated deformed scans derived from a single volume ($\mathbf{V}_{9,1}$ from $\mathcal{P}_9$).
All trained networks were subsequently used to register simulated deformed scans from the independent test set back onto their original volumes (Fig.~\ref{fig:ResSimDefs}).
In all cross-validation rounds, the incremental training strategy resulted in better registration performance compared to training on a single volume, with mean DSC values of $0.929 \pm 0.037$ and $0.883 \pm 0.033$, and mean SSIM values of $0.984 \pm 0.009$ and $0.969 \pm 0.007$, respectively.
The deformable registrations performed using NiftyReg resulted in the best registration results, with a mean DSC of $0.988 \pm 0.003$ and a mean SSIM of $0.995 \pm 0.002$, although it should be noted that this registration method was about $300$ times slower than one forward pass through the VoxelMorph framework (Table~\ref{table:computPerform}).

To evaluate the impact of the inter- and intra-patient variations on the longitudinal abdominopelvic CT dataset, $\mathcal{D}$, on the registration performance, all trained networks were also used to register real scan pairs, i.e., mapping sequential time-points back onto the reference scan (time-point $0$).
Fig.~\ref{fig:ResRealDefs} shows the DSC and SSIM values between the real scan pairs before registration, after registration using the VoxelMorph framework trained on single volume or incrementally, and NiftyReg.
The differences between the scan pairs before registration greatly varied between patients, with DSC and SSIM values ranging from $0.567$ to $0.920$ and from $0.693$ and $0.918$, respectively.
Although the VoxelMorph networks were trained using only simulated deformations, the incrementally trained networks improved the DSC between the real scan pairs for $6$ out of the $12$ patients, whereas the network trained on a single volume improved the DSC for $4$ out of the $12$ patients (Fig.~\ref{fig:ResRealDefs}).
Furthermore, all VoxelMorph-based models improved the SSIM between the real scan pairs for all patients except patient $\mathcal{P}_6$.
However, it should be noted that none of the networks trained in this study achieved registration results comparable to NiftyReg.

\begin{figure}[!ht]
    \centering
    \includegraphics[width=0.7\textwidth]{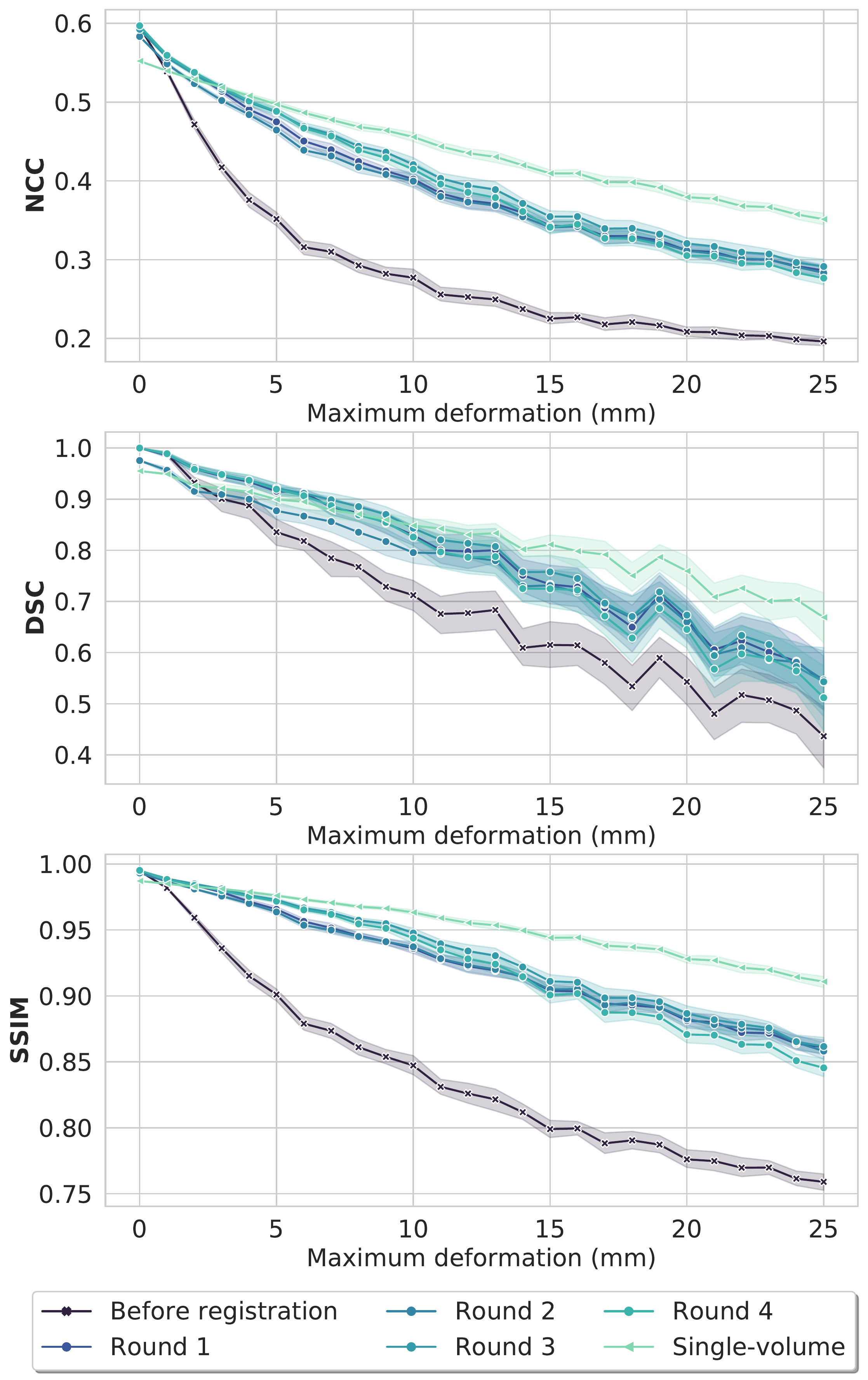}
    \vspace{-0.4cm}
    \caption{Registration performance on increasingly large displacements in terms of NCC, DSC, and SSIM.}
    \label{fig:MaxDef}
\end{figure}

\subsubsection{Large displacements}
\label{sec:largeDispl}

In addition to variations between patients, mapping large displacements may also form a challenge for deep learning based deformable registration methods.
In order to evaluate the effect of the size of the displacements on the registration performance of the networks trained in this study, an additional test set was created by simulating $K$ DVFs $\phi_k$ ($k=1,\ldots,K$) with maximum displacements ranging from 0 mm (i.e., no deformation) to $25$ mm (i.e., structures moving across the entire abdominal and pelvic regions) in steps of $1$ mm, with $K = 30$ in each step. These DVFs were used to deform the same volume that was used to generate the training data to train the single-volume network ($\mathbf{V}_{9,1}$ from $\mathcal{P}_9$), after which the deformed images were mapped back onto the original volume using the trained VoxelMorph networks and NiftyReg.

Fig.~\ref{fig:MaxDef} shows the mean NCC (see Eq.(~\ref{eq:NCC})), DSC, and SSIM values for the full range of maximum displacements.
The network trained on a single-volume ideally represents the ``best possible'' (although clinically unrealistic) scenario in which the network was trained and tested on the same volume.
This network thus performed better on larger displacements, whereas the incrementally trained networks performed better for small deformations up to $5$ mm.

\section{Discussion and conclusions}
\label{sec:discussion}

In recent years, an increasing number of studies have focused on using deep neural networks for deformable image registration because such methods offer fast or nearly real-time registration~\cite{yang2017,deVos2019DLIR,balakrishnan2019,sokooti2017,eppenhof2019,eppenhof2019TMI,hu2018}.
However, their application to abdominopelvic CT images remains limited because of the large intra- and inter-patient variations, the not fully diffeomorphic nature of the deformations, and the limited availability of large numbers of well-annotated images for training.

In the present study, we demonstrated that removing extracorporeal structures aids deformable registration of abdominopelvic CT images when using both traditional and deep learning approaches.
Along with the registration of multiple CT scans over time, in which the table design and shape may differ and affect the registration process, the devised method based on region-growing~\cite{rundoMBEC2016} could also be valuable for multimodal image registration tasks because the scanner table is not visible on MRI and PET \cite{zhu2012}.
Another practical use case could be radiation treatment-planning, in which the CT table influences the dose distribution since the table used during imaging typically has different beam attenuation characteristics compared to the treatment table \cite{spezi2008}.

To address the remaining challenges of our abdominopelvic CT dataset, we generated training data for our network by synthetically deforming the CT images.
Such synthetically deformed images can be employed for different purposes: (\textit{i}) training a neural network for deformable image registration on a relatively small clinical dataset; and (\textit{ii}) evaluation, e.g., testing the ability of a network to register increasingly large displacements.
Simulated DVFs have already been successfully used for supervised learning of deformable image registration \cite{sokooti2017,eppenhof2019TMI}.
As a future development, we plan to introduce an additional penalty term into the loss function of our registration method to exploit the known simulated DVFs during training, which would allow the training process to gradually transition from semi-supervised to unsupervised learning.

To exploit the longitudinal nature of our dataset and enable training on small amounts of training data, we propose a novel incremental strategy to train the VoxelMorph framework.
Incremental learning has shown potential for image classification \cite{castro2018} and medical image segmentation \cite{BenNaceur2018}, although the so-called catastrophic forgetting \cite{lee2017} still remains a challenge.
The incremental training of neural networks for longitudinal image registration could, therefore, benefit from introducing a penalty term into the loss function to balance the registration performance on new images while minimizing forgetting of previous images.

In the long term, parameter-efficient, single-shot deep learning solutions for deformable image registration would enable the development of novel end-to-end approaches, such as task-adapted image reconstruction \cite{adler2018}.
From a clinical perspective, automated registration of longitudinal imaging data is a prerequisite for exploiting the full potential of standard-of-care CT images for treatment response assessment in patients with bone metastases. 
A successful approach might find potential applications in the most frequently occurring malignancies that have a tendency to metastasize to bone, i.e., prostate, lung, and breast cancer \cite{macedo2017}.

\section*{Acknowledgements}

MvE acknowledges financial support from the Netherlands Organisation for
Scientific Research (NWO) [project number 639.073.506] and the Royal Dutch Academy of Sciences (KNAW) [Van Leersum Grant 2018].

This work was also partially supported by The Mark Foundation for Cancer Research and Cancer Research UK Cambridge Centre [C9685/A25177].
Additional support was also provided by the National Institute of Health Research (NIHR) Cambridge Biomedical Research Centre.
The views expressed are those of the authors and not necessarily those of the NHS, the NIHR, or the Department of Health and Social Care.

CBS acknowledges support from the Leverhulme Trust project on `Breaking the non-convexity barrier', EPSRC grant Nr. EP/M00483X/1, EP/S026045/1, the EPSRC Centre Nr. EP/N014588/1, the RISE projects CHiPS and NoMADS, the Cantab Capital Institute for the Mathematics of Information and the Alan Turing Institute.

Microsoft Radiomics was provided to the Addenbrooke's Hospital (Cambridge University Hospitals NHS Foundation Trust, Cambridge, UK) by the Microsoft InnerEye project.

\section*{References}

\bibliography{mybibfile}

\begin{thebibliography}{10}
\expandafter\ifx\csname url\endcsname\relax
  \def\url#1{\texttt{#1}}\fi
\expandafter\ifx\csname urlprefix\endcsname\relax\def\urlprefix{URL }\fi
\expandafter\ifx\csname href\endcsname\relax
  \def\href#1#2{#2} \def\path#1{#1}\fi

\bibitem{pluim2000}
J.~P.~W. Pluim, J.~B.~A. Maintz, M.~A. Viergever, Image registration by
  maximization of combined mutual information and gradient information, IEEE
  Trans. Med. Imaging 19~(8) (2000) 809--814.
\newblock \href {http://dx.doi.org/10.1109/42.876307}
  {\path{doi:10.1109/42.876307}}.

\bibitem{sokooti2017}
H.~Sokooti, B.~de~Vos, F.~Berendsen, B.~P. Lelieveldt, I.~I{\v{s}}gum,
  M.~Staring, Nonrigid image registration using multi-scale {3D} convolutional
  neural networks, in: Proc. International Conference on Medical Image
  Computing and Computer-Assisted Intervention (MICCAI), Springer, 2017, pp.
  232--239.
\newblock \href {http://dx.doi.org/10.1007/978-3-319-66182-7 27}
  {\path{doi:10.1007/978-3-319-66182-7 27}}.

\bibitem{litjens2017survey}
G.~Litjens, T.~Kooi, B.~E. Bejnordi, A.~A.~A. Setio, F.~Ciompi, M.~Ghafoorian,
  J.~A. Van Der~Laak, B.~Van~Ginneken, C.~I. S{\'a}nchez, A survey on deep
  learning in medical image analysis, Med. Image Anal. 42 (2017) 60--88.
\newblock \href {http://dx.doi.org/10.1109/TMI.2014.2303821}
  {\path{doi:10.1109/TMI.2014.2303821}}.

\bibitem{shen2017}
D.~Shen, G.~Wu, H.-I. Suk, Deep learning in medical image analysis, Annu. Rev.
  Biomed. Eng. 19 (2017) 221--248.
\newblock \href {http://dx.doi.org/10.1146/annurev-bioeng-071516-044442}
  {\path{doi:10.1146/annurev-bioeng-071516-044442}}.

\bibitem{haskins2020}
G.~Haskins, U.~Kruger, P.~Yan, Deep learning in medical image registration: a
  survey, Mach. Vis. Appl. 31~(1) (2020) 8.
\newblock \href {http://dx.doi.org/10.1007/s00138-020-01060-x}
  {\path{doi:10.1007/s00138-020-01060-x}}.

\bibitem{han2019CIKM}
C.~Han, K.~Murao, T.~Noguchi, Y.~Kawata, F.~Uchiyama, L.~Rundo, H.~Nakayama,
  S.~Satoh, Learning more with less: {Conditional} {PGGAN}-based data
  augmentation for brain metastases detection using highly-rough annotation on
  {MR} images, in: Proc. ACM International Conference on Information and
  Knowledge Management (CIKM), 2019, pp. 119--127.
\newblock \href {http://dx.doi.org/10.1145/3357384.3357890}
  {\path{doi:10.1145/3357384.3357890}}.

\bibitem{xu2016}
Z.~Xu, C.~P. Lee, M.~P. Heinrich, M.~Modat, D.~Rueckert, S.~Ourselin, R.~G.
  Abramson, B.~A. Landman, Evaluation of six registration methods for the human
  abdomen on clinically acquired {CT}, IEEE Trans. Biomed. Eng. 63~(8) (2016)
  1563--1572.
\newblock \href {http://dx.doi.org/10.1109/TBME.2016.2574816}
  {\path{doi:10.1109/TBME.2016.2574816}}.

\bibitem{freiman2011}
M.~Freiman, S.~D. Voss, S.~K. Warfield, Abdominal images non-rigid registration
  using local-affine diffeomorphic demons, in: Proc. International MICCAI
  Workshop on Computational and Clinical Challenges in Abdominal Imaging,
  Springer, 2011, pp. 116--124.
\newblock \href {http://dx.doi.org/10.1007/978-3-642-28557-8_15}
  {\path{doi:10.1007/978-3-642-28557-8_15}}.

\bibitem{yankeelov2016}
T.~E. Yankeelov, D.~A. Mankoff, L.~H. Schwartz, F.~S. Lieberman, J.~M. Buatti,
  J.~M. Mountz, et~al., Quantitative imaging in cancer clinical trials, Clin.
  Cancer Res. 22~(2) (2016) 284--290.
\newblock \href {http://dx.doi.org/10.1158/1078-0432.CCR-14-3336}
  {\path{doi:10.1158/1078-0432.CCR-14-3336}}.

\bibitem{blackledge2014}
M.~D. Blackledge, D.~J. Collins, N.~Tunariu, M.~R. Orton, A.~R. Padhani, M.~O.
  Leach, D.-M. Koh, Assessment of treatment response by total tumor volume and
  global apparent diffusion coefficient using diffusion-weighted {MRI} in
  patients with metastatic bone disease: a feasibility study, PLoS One 9~(4).
\newblock \href {http://dx.doi.org/10.1371/journal.pone.0091779}
  {\path{doi:10.1371/journal.pone.0091779}}.

\bibitem{reischauer2018}
C.~Reischauer, R.~Patzwahl, D.-M. Koh, J.~M. Froehlich, A.~Gutzeit, Texture
  analysis of apparent diffusion coefficient maps for treatment response
  assessment in prostate cancer bone metastases -- a pilot study, Eur. J.
  Radiol. 101 (2018) 184--190.
\newblock \href {http://dx.doi.org/10.1016/j.ejrad.2018.02.024}
  {\path{doi:10.1016/j.ejrad.2018.02.024}}.

\bibitem{balakrishnan2019}
G.~Balakrishnan, A.~Zhao, M.~R. Sabuncu, J.~Guttag, A.~V. Dalca, {VoxelMorph}:
  a learning framework for deformable medical image registration, IEEE Trans.
  Med. Imaging\href {http://dx.doi.org/10.1109/TMI.2019.2897538}
  {\path{doi:10.1109/TMI.2019.2897538}}.

\bibitem{modat2010}
M.~Modat, G.~R. Ridgway, Z.~A. Taylor, M.~Lehmann, J.~Barnes, D.~J. Hawkes,
  N.~C. Fox, S.~Ourselin, Fast free-form deformation using graphics processing
  units, Comput. Methods Programs Biomed. 98~(3) (2010) 278--284.
\newblock \href {http://dx.doi.org/10.1016/j.cmpb.2009.09.002}
  {\path{doi:10.1016/j.cmpb.2009.09.002}}.

\bibitem{lee2015evaluation}
C.~P. Lee, Z.~Xu, R.~P. Burke, R.~Baucom, B.~K. Poulose, R.~G. Abramson, B.~A.
  Landman, Evaluation of five image registration tools for abdominal {CT}:
  pitfalls and opportunities with soft anatomy, in: Medical Imaging 2015: Image
  Processing, Vol. 9413, International Society for Optics and Photonics, 2015,
  p. 94131N.

\bibitem{klein2007evaluation}
S.~Klein, M.~Staring, J.~P. Pluim, Evaluation of optimization methods for
  nonrigid medical image registration using mutual information and {B-splines},
  IEEE Trans. Image Process. 16~(12) (2007) 2879--2890.
\newblock \href {http://dx.doi.org/10.1109/TIP.2007.909412}
  {\path{doi:10.1109/TIP.2007.909412}}.

\bibitem{bernon2001}
J.~Bernon, V.~Boudousq, J.~Rohmer, M.~Fourcade, M.~Zanca, M.~Rossi,
  D.~Mariano-Goulart, A comparative study of {Powell's and Downhill Simplex}
  algorithms for a fast multimodal surface matching in brain imaging, Comput.
  Med. Imaging Graph. 25~(4) (2001) 287--297.
\newblock \href {http://dx.doi.org/10.1016/S0895-6111(00)00073-2}
  {\path{doi:10.1016/S0895-6111(00)00073-2}}.

\bibitem{thirion1998}
J.~Thirion, Image matching as a diffusion process: an analogy with {Maxwell}'s
  demons, Med. Image Anal. 2~(3) (1998) 243--260.
\newblock \href {http://dx.doi.org/10.1016/S1361-8415(98)80022-4}
  {\path{doi:10.1016/S1361-8415(98)80022-4}}.

\bibitem{vercauteren2009}
T.~Vercauteren, X.~Pennec, A.~Perchant, N.~Ayache, Diffeomorphic demons:
  efficient non-parametric image registration, NeuroImage 45~(1) (2009)
  S61--S72.
\newblock \href {http://dx.doi.org/10.1016/j.neuroimage.2008.10.040}
  {\path{doi:10.1016/j.neuroimage.2008.10.040}}.

\bibitem{rundo2016SSCI}
L.~Rundo, A.~Tangherloni, C.~Militello, M.~C. Gilardi, G.~Mauri, Multimodal
  medical image registration using particle swarm optimization: a review, in:
  Proc. Symposium Series on Computational Intelligence (SSCI), IEEE, 2016, pp.
  1--8.
\newblock \href {http://dx.doi.org/10.1109/SSCI.2016.7850261}
  {\path{doi:10.1109/SSCI.2016.7850261}}.

\bibitem{klein2009elastix}
S.~Klein, M.~Staring, K.~Murphy, M.~A. Viergever, J.~P. Pluim, {elastix}: a
  toolbox for intensity-based medical image registration, IEEE Trans. Med.
  Imaging 29~(1) (2009) 196--205.
\newblock \href {http://dx.doi.org/10.1109/TMI.2009.2035616}
  {\path{doi:10.1109/TMI.2009.2035616}}.

\bibitem{tustison2014}
N.~J. Tustison, P.~A. Cook, A.~Klein, G.~Song, S.~R. Das, J.~T. Duda, et~al.,
  Large-scale evaluation of {ANTs} and {FreeSurfer} cortical thickness
  measurements, NeuroImage 99 (2014) 166--179.
\newblock \href {http://dx.doi.org/10.1016/j.neuroimage.2014.05.044}
  {\path{doi:10.1016/j.neuroimage.2014.05.044}}.

\bibitem{modersitzki2009FAIR}
J.~Modersitzki, FAIR: Flexible Algorithms for Image Registration, Vol.~6,
  Society for Industrial and Applied Mathematics (SIAM), Philadelphia, PA, USA,
  2009.

\bibitem{wu2013}
G.~Wu, M.~Kim, Q.~Wang, Y.~Gao, S.~Liao, D.~Shen, Unsupervised deep feature
  learning for deformable registration of {MR} brain images, in: Proc.
  International Conference on Medical Image Computing and Computer-Assisted
  Intervention (MICCAI), Vol. 8150 of LNCS, Springer, 2013, pp. 649--656.
\newblock \href {http://dx.doi.org/10.1007/978-3-642-40763-5_80}
  {\path{doi:10.1007/978-3-642-40763-5_80}}.

\bibitem{wu2016}
G.~Wu, M.~Kim, Q.~Wang, B.~C. Munsell, D.~Shen, Scalable high-performance image
  registration framework by unsupervised deep feature representations learning,
  IEEE Trans. Biomed. Eng. 63~(7) (2016) 1505--1516.
\newblock \href {http://dx.doi.org/10.1109/TBME.2015.2496253}
  {\path{doi:10.1109/TBME.2015.2496253}}.

\bibitem{Cheng2018}
X.~Cheng, L.~Zhang, Y.~Zheng, Deep similarity learning for multimodal medical
  images, Comput. Methods Biomech. Biomed. Eng. Imaging Vis. 6~(3) (2018)
  248--252.
\newblock \href {http://dx.doi.org/10.1080/21681163.2015.1135299}
  {\path{doi:10.1080/21681163.2015.1135299}}.

\bibitem{yang2017}
X.~Yang, R.~Kwitt, M.~Styner, M.~Niethammer, Quicksilver: Fast predictive image
  registration--a deep learning approach, NeuroImage 158 (2017) 378--396.

\bibitem{eppenhof2019TMI}
K.~A. Eppenhof, J.~P. Pluim, Pulmonary {CT} registration through supervised
  learning with convolutional neural networks, IEEE Trans. Med. Imaging 38~(5)
  (2019) 1097--1105.
\newblock \href {http://dx.doi.org/10.1109/TMI.2018.2878316}
  {\path{doi:10.1109/TMI.2018.2878316}}.

\bibitem{Ha2019MIDL}
I.~Y. Ha, L.~Hansen, M.~Wilms, M.~P. Heinrich,
  \href{https://openreview.net/forum?id=BJg6ntQRtE}{Geometric deep learning and
  heatmap prediction for large deformation registration of abdominal and
  thoracic {CT}}, in: International Conference on Medical Imaging with Deep
  Learning (MIDL), London, United Kingdom, 2019, pp. 1--5.
\newline\urlprefix\url{https://openreview.net/forum?id=BJg6ntQRtE}

\bibitem{hu2018}
Y.~Hu, M.~Modat, E.~Gibson, W.~Li, N.~Ghavami, E.~Bonmati, G.~Wang, S.~Bandula,
  C.~M. Moore, M.~Emberton, et~al., Weakly-supervised convolutional neural
  networks for multimodal image registration, Med. Image Anal. 49 (2018) 1--13.
\newblock \href {http://dx.doi.org/10.1016/j.media.2018.07.002}
  {\path{doi:10.1016/j.media.2018.07.002}}.

\bibitem{hu2018adversarial}
Y.~Hu, E.~Gibson, N.~Ghavami, E.~Bonmati, C.~M. Moore, M.~Emberton,
  T.~Vercauteren, J.~A. Noble, D.~C. Barratt, Adversarial deformation
  regularization for training image registration neural networks, in: Proc.
  International Conference on Medical Image Computing and Computer-Assisted
  Intervention (MICCAI), Vol. 11070 of LNCS, Springer, 2018, pp. 774--782.
\newblock \href {http://dx.doi.org/10.1007/978-3-030-00928-1_87}
  {\path{doi:10.1007/978-3-030-00928-1_87}}.

\bibitem{yan2018adversarial}
P.~Yan, S.~Xu, A.~R. Rastinehad, B.~J. Wood, Adversarial image registration
  with application for {MR} and {TRUS} image fusion, in: Proc. International
  Workshop on Machine Learning in Medical Imaging (MLMI), Vol. 11046 of LNCS,
  Springer, 2018, pp. 197--204.
\newblock \href {http://dx.doi.org/10.1007/978-3-030-00919-9_23}
  {\path{doi:10.1007/978-3-030-00919-9_23}}.

\bibitem{tanner2018}
C.~Tanner, F.~Ozdemir, R.~Profanter, V.~Vishnevsky, E.~Konukoglu, O.~Goksel,
  Generative adversarial networks for {MR-CT} deformable image registration,
  arXiv preprint arXiv:1807.07349.

\bibitem{krebs2019}
J.~Krebs, H.~Delingette, B.~Mailh{\'e}, N.~Ayache, T.~Mansi, Learning a
  probabilistic model for diffeomorphic registration, IEEE Trans. Med. Imaging
  38.
\newblock \href {http://dx.doi.org/10.1109/TMI.2019.2897112}
  {\path{doi:10.1109/TMI.2019.2897112}}.

\bibitem{jaderberg2015}
M.~Jaderberg, K.~Simonyan, A.~Zisserman, et~al., Spatial transformer networks,
  in: Proc. Advances in Neural Information Processing Systems (NIPS), 2015, pp.
  2017--2025.

\bibitem{deVos2019DLIR}
B.~D. de~Vos, F.~F. Berendsen, M.~A. Viergever, H.~Sokooti, M.~Staring,
  I.~I{\v{s}}gum, A deep learning framework for unsupervised affine and
  deformable image registration, Med. Image Anal. 52 (2019) 128--143.
\newblock \href {http://dx.doi.org//10.1016/j.media.2018.11.010}
  {\path{doi:/10.1016/j.media.2018.11.010}}.

\bibitem{ronneberger2015}
O.~Ronneberger, P.~Fischer, T.~Brox, {U-Net}: Convolutional networks for
  biomedical image segmentation, in: Proc. Conference on Medical Image
  Computing and Computer-Assisted Intervention (MICCAI), Vol. 9351 of LNCS,
  Springer, 2015, pp. 234--241.
\newblock \href {http://dx.doi.org/10.1007/978-3-319-24574-4_28}
  {\path{doi:10.1007/978-3-319-24574-4_28}}.

\bibitem{eppenhof2019}
K.~A. Eppenhof, M.~W. Lafarge, J.~P. Pluim, Progressively growing convolutional
  networks for end-to-end deformable image registration, in: Medical Imaging
  2019: Image Processing, Vol. 10949, International Society for Optics and
  Photonics, 2019, p. 109491C.
\newblock \href {http://dx.doi.org/10.1117/12.2512428}
  {\path{doi:10.1117/12.2512428}}.

\bibitem{kim2019unsupervised}
B.~Kim, J.~Kim, J.-G. Lee, D.~H. Kim, S.~H. Park, J.~C. Ye, Unsupervised
  deformable image registration using cycle-consistent {CNN}, in: Proc.
  International Conference on Medical Image Computing and Computer-Assisted
  Intervention (MICCAI), Vol. 11769 of LNCS, Springer, 2019, pp. 166--174.
\newblock \href {http://dx.doi.org/10.1007/978-3-030-32226-7_19}
  {\path{doi:10.1007/978-3-030-32226-7_19}}.

\bibitem{zhu2017cycle}
J.-Y. Zhu, T.~Park, P.~Isola, A.~A. Efros, Unpaired image-to-image translation
  using cycle-consistent adversarial networks, in: Proc. IEEE International
  Conference on Computer Vision (ICCV), 2017, pp. 2223--2232.
\newblock \href {http://dx.doi.org/10.1109/ICCV.2017.244}
  {\path{doi:10.1109/ICCV.2017.244}}.

\bibitem{dosovitskiy2015}
A.~Dosovitskiy, P.~Fischer, E.~Ilg, P.~Hausser, C.~Hazirbas, V.~Golkov, P.~Van
  Der~Smagt, D.~Cremers, T.~Brox, {FlowNet}: learning optical flow with
  convolutional networks, in: Proc. Conference on Computer Vision and Pattern
  Recognition (CVPR), IEEE, 2015, pp. 2758--2766.
\newblock \href {http://dx.doi.org/10.1109/ICCV.2015.316}
  {\path{doi:10.1109/ICCV.2015.316}}.

\bibitem{li2019}
T.~Li, M.~Zhang, W.~Qi, E.~Asma, J.~Qi, Motion correction of respiratory-gated
  {PET} image using deep learning based image registration framework, in: Proc.
  15th International Meeting on Fully Three-Dimensional Image Reconstruction in
  Radiology and Nuclear Medicine, Vol. 11072, International Society for Optics
  and Photonics, 2019, p. 110720Q.
\newblock \href {http://dx.doi.org/10.1117/12.2534851}
  {\path{doi:10.1117/12.2534851}}.

\bibitem{burns2013}
J.~E. Burns, J.~Yao, T.~S. Wiese, H.~E. Mu{\~n}oz, E.~C. Jones, R.~M. Summers,
  Automated detection of sclerotic metastases in the thoracolumbar spine at
  {CT}, Radiology 268~(1) (2013) 69--78.
\newblock \href {http://dx.doi.org/10.1148/radiol.13121351}
  {\path{doi:10.1148/radiol.13121351}}.

\bibitem{mouliere2018}
F.~Mouliere, D.~Chandrananda, A.~M. Piskorz, E.~K. Moore, J.~Morris, L.~B.
  Ahlborn, et~al., Enhanced detection of circulating tumor {DNA} by fragment
  size analysis, Sci. Trans. Med. 10~(466) (2018) eaat4921.
\newblock \href {http://dx.doi.org/10.1126/scitranslmed.aat4921}
  {\path{doi:10.1126/scitranslmed.aat4921}}.

\bibitem{gambino2011}
O.~Gambino, E.~Daidone, M.~Sciortino, R.~Pirrone, E.~Ardizzone, Automatic skull
  stripping in {MRI} based on morphological filters and fuzzy c-means
  segmentation, in: Proc. Annual International Conference of the IEEE
  Engineering in Medicine and Biology Society (EMBS), IEEE, 2011, pp.
  5040--5043.
\newblock \href {http://dx.doi.org/10.1109/IEMBS.2011.6091248}
  {\path{doi:10.1109/IEMBS.2011.6091248}}.

\bibitem{rundoMBEC2016}
L.~Rundo, C.~Militello, S.~Vitabile, C.~Casarino, G.~Russo, M.~Midiri, M.~C.
  Gilardi, Combining split-and-merge and multi-seed region growing algorithms
  for uterine fibroid segmentation in {MRgFUS} treatments, Med. Biol. Eng.
  Comput. 54~(7) (2016) 1071--1084.
\newblock \href {http://dx.doi.org/10.1007/s11517-015-1404-6}
  {\path{doi:10.1007/s11517-015-1404-6}}.

\bibitem{avants2008}
B.~B. Avants, C.~L. Epstein, M.~Grossman, J.~C. Gee, Symmetric diffeomorphic
  image registration with cross-correlation: evaluating automated labeling of
  elderly and neurodegenerative brain, Med. Image Anal. 12~(1) (2008) 26--41.
\newblock \href {http://dx.doi.org/10.1016/j.media.2007.06.004}
  {\path{doi:10.1016/j.media.2007.06.004}}.

\bibitem{chefrour2019}
A.~Chefrour, Incremental supervised learning: algorithms and applications in
  pattern recognition, Evol. Intell. 12 (2019) 97--112.
\newblock \href {http://dx.doi.org/10.1007/s12065-019-00203-y}
  {\path{doi:10.1007/s12065-019-00203-y}}.

\bibitem{goyal2017}
P.~Goyal, P.~Doll{\'a}r, R.~Girshick, P.~Noordhuis, L.~Wesolowski, A.~Kyrola,
  A.~Tulloch, Y.~Jia, K.~He, Accurate, large minibatch {SGD}: training
  {ImageNet} in 1 hour, arXiv preprint arXiv:1706.02677.

\bibitem{wang2004}
Z.~Wang, A.~C. Bovik, H.~R. Sheikh, E.~P. Simoncelli, Image quality assessment:
  from error visibility to structural similarity, IEEE Trans. Image Process.
  13~(4) (2004) 600--612.
\newblock \href {http://dx.doi.org/10.1109/TIP.2003.819861}
  {\path{doi:10.1109/TIP.2003.819861}}.

\bibitem{wilcoxon1992}
F.~Wilcoxon, Individual comparisons by ranking methods, Biometrics Bull. 1~(6)
  (80--83) 196--202.
\newblock \href {http://dx.doi.org/10.2307/3001968}
  {\path{doi:10.2307/3001968}}.

\bibitem{zhu2012}
Y.-M. Zhu, S.~M. Cochoff, R.~Sukalac, Automatic patient table removal in {CT}
  images, J. Digit. Imaging 25~(4) (2012) 480--485.
\newblock \href {http://dx.doi.org/10.1007/s10278-012-9454-x}
  {\path{doi:10.1007/s10278-012-9454-x}}.

\bibitem{spezi2008}
E.~Spezi, A.~L. Angelini, F.~Romani, A.~Guido, F.~Bunkheila, M.~Ntreta,
  A.~Ferri, Evaluating the influence of the {Siemens IGRT} carbon fibre
  tabletop in head and neck {IMRT}, Radiotherapy and Oncology 89~(1) (2008)
  114--122.
\newblock \href {http://dx.doi.org/10.1016/j.radonc.2008.06.011}
  {\path{doi:10.1016/j.radonc.2008.06.011}}.

\bibitem{castro2018}
F.~M. Castro, M.~J. Mar{\'\i}n-Jim{\'e}nez, N.~Guil, C.~Schmid, K.~Alahari,
  End-to-end incremental learning, in: Proc. European Conference on Computer
  Vision (ECCV), Vol. 11216 of LNCS, Springer, 2018, pp. 233--248.
\newblock \href {http://dx.doi.org/10.1007/978-3-030-01258-8_15}
  {\path{doi:10.1007/978-3-030-01258-8_15}}.

\bibitem{BenNaceur2018}
M.~Ben~naceur, R.~Saouli, M.~Akil, R.~Kachouri, Fully automatic brain tumor
  segmentation using end-to-end incremental deep neural networks in {MRI}
  images, Comput. Methods Programs Biomed. 166 (2018) 39--49.
\newblock \href {http://dx.doi.org/10.1016/j.cmpb.2018.09.007}
  {\path{doi:10.1016/j.cmpb.2018.09.007}}.

\bibitem{lee2017}
S.-W. Lee, J.-H. Kim, J.~Jun, J.-W. Ha, B.-T. Zhang, Overcoming catastrophic
  forgetting by incremental moment matching, in: Proc. Advances in Neural
  Information Processing Systems (NIPS), 2017, pp. 4652--4662.

\bibitem{adler2018}
J.~Adler, S.~Lunz, O.~Verdier, C.-B. Sch{\"o}nlieb, O.~{\"O}ktem, Task adapted
  reconstruction for inverse problems, arXiv preprint arXiv:1809.00948.

\bibitem{macedo2017}
F.~Macedo, K.~Ladeira, F.~Pinho, N.~Saraiva, N.~Bonito, L.~Pinto,
  F.~Gon{\c{c}}alves, Bone metastases: an overview, Oncol. Rev. 11~(1).
\newblock \href {http://dx.doi.org/10.4081/oncol.2017.321}
  {\path{doi:10.4081/oncol.2017.321}}.

\end{thebibliography}

\end{document}